\documentclass[reprint,amsmath,amssymb,braket,aps,prb,longbibliography]{revtex4-1} 
\usepackage{graphicx,color,braket,amsmath,dcolumn,bm, float}
\usepackage[colorlinks=true,linkcolor=blue]{hyperref}

\begin{document}
\title{Electromagnetic diffraction and bidirectional plasmon launching in partially gated 2d systems}

\author{Ilia Moiseenko}
\affiliation{Laboratory of 2d Materials for Optoelectonics, Moscow Institute of Physics and Technology, Dolgoprudny 141700, Russia}

\author{Egor Nikulin}
\affiliation{Russian Quantum Center, Skolkovo, 143025 Moscow, Russia}

\author{Dmitry Svintsov}
\email{svintcov.da@mipt.ru}
\affiliation{Laboratory of 2d Materials for Optoelectonics, Moscow Institute of Physics and Technology, Dolgoprudny 141700, Russia}

\begin{abstract}
Partially gated two-dimensional electron systems (2DES) represent the basic building block of prospective optoelectronic devices, including electromagnetic detectors and sources. At the same time, the electrodynamic properties of such structures have been addressed only with numerical simulations. Here, we provide an exact solution of electromagnetic scattering problem at a partially gated 2DES using the Wiener-Hopf technique. We find that incident $p$-polarized field is enhanced in immediate vicinity of gate edge. \textcolor{black}{The edge acts as a plasmonic coupler that launches bidirectional (gated and ungated) plasmons in weakly-dissipative 2DES with impedance of inductive type. Electric fields in these waves markedly exceeds the incident field, especially in the limit of small gate-2DES separation. The amplitude of the ungated wave exceeds that of gated for 2DES reactance above the free-space impedance. Both amplitudes are maximized for the 2DES reactance of the order of free space impedance timed by the ratio of free space wavelength and gate-2DES separation.} 

\end{abstract}
\maketitle

\section{Introduction}

Plasmonics represents a physical paradigm for concentrating and manipulating the electromagnetic energy at the deep subwavelength scale. It exploits the electromagnetic waves supported by conductive media, plasmons. Among various subtypes of plasmons, those supported by two-dimensional (2d) electron systems~\cite{Stern1967,Allen1977}and  graphene~\cite{Ryzhii2007a,Woessner2015} are especially attractive. The principal reason is their ultra-slow velocity, about two orders below the speed of light at typical experimental conditions, and consequently -- strong confinement. The 2d plasmon velocity depends on the strength of Coulomb interactions between charges, and goes down with introduction of screening. In 2d systems with metallic gates, the screening image charges result in linear (sound-like) plasmon dispersion~\cite{chaplik1972possible}, where the wave velocity takes quite low values limited only by the thermal or Fermi velocity of individual charge carriers~\cite{Heitmann_Fermi_pressure,Lundeberg2017c,Ryzhii2007a}.

Strong confinement of gated 2d plasmons suggests that they can efficiently mediate the interaction between charge carriers and electromagnetic fields. In particular, excitation of gated plasmoms in 2d diodes and transistors can result in their elevated photoresponsivity~\cite{Muravev_detector,Woessner_electrical,Castilla_electrical_spectroscopy}. Tuning of plasmon velocity with gate voltage further enables on-chip radiation spectroscopy with these devices, particularly, in the terahertz and sub-terahertz ranges~\cite{Bandurin_resonant,Muravev_interferometer}. Sensitivity of 2d plasmon resonance frequecny to the tiny variations in electromagnetic environment is the basis of plasmon-enhanced biosensing~\cite{Rodrigo_biosensing,Bylinkin2024}. Further, 2d plasmons can be excited -- either thermally~\cite{Graphene_thermal_emission} or coherently~\cite{Dyakonov1993,ElFatimy2010} -- under passage of direct current in 2DES. Radiative decay of these excitations forms the basis of plasmonic radiation sources.

\textcolor{black}{Despite the numerous applications of ultra-confined 2d plasmons, their theory so far has been largely limited to unbounded 2d systems, where spectral properties have been studied extensively~\cite{Stern1967,chaplik1972possible,Hwang2007a,Ryzhii2007a}. The theory of 2d plasmon excitation in bounded 2DES is still on the early stage of development.} Excitation of 2d plasmons by free space electromagnetic waves is impeded by momentum mismatch. The free-space wave should be scattered at some obstacle (isolated~\cite{Limits_to_plasmon_coupling} or periodic~\cite{Chaplik_emission_absorption,Fateev_transormation}) with formation of highly non-uniform local field. This non-uniform field, in turn, can couple to the 2d plasmons. From analytical viewpoint, solution of Maxwell's equations in the presence of both 2d conductors and obstacles is possible in very limited cases, the most known being the polarizable dipole above the 2DES~\cite{McLeod2014}. From computational viewpoint, solution of scattering problems for 2d plasmons requires very large (exceeding $\lambda_0$) and very dense (below $\lambda_{\rm pl}$) computational grids, which leads to long simulation times.

\begin{figure}
\center{\includegraphics[width=0.9\linewidth]{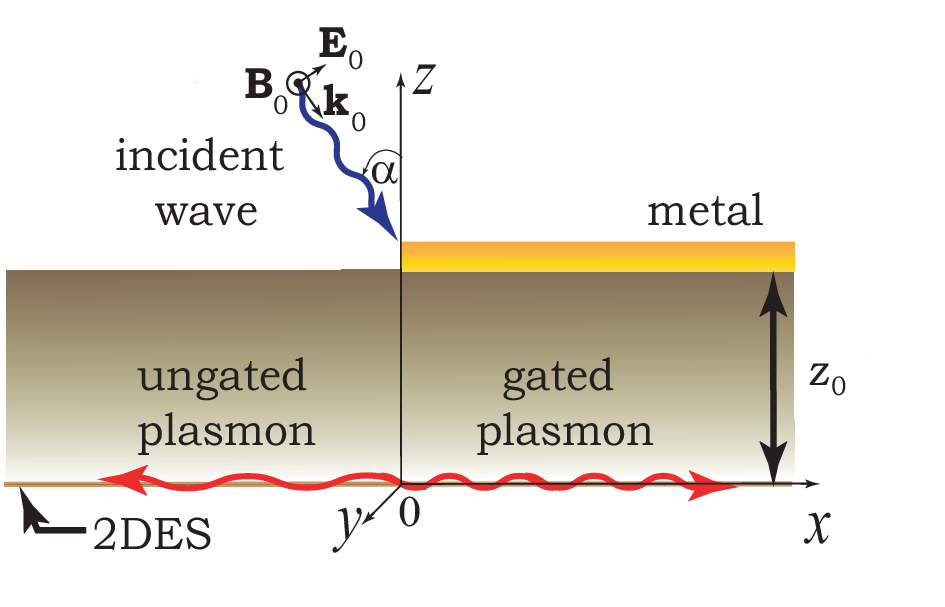}}
\caption{Schematic view of the studied structure: electromagnetic wave incident at an arbitrary angle $\alpha$ on a partly gated 2DES. Red wavy lines show the gated and ungated plasmons running away from the junction}
\label{structure}
\end{figure}

In this paper, we obtain an exact solution for the electromagnetic diffraction and launching of 2d plasmons at the edge of the metal gate covering \textcolor{black}{half of} the two-dimensional electron system with uniform conductivity $\sigma_{\rm 2d}$ \textcolor{black}{. In the following, such structure will be referred to as ''partially gated 2DES''}. The setup under study is shown in Fig.~\ref{structure}. Apart from being a coupler between free-space photons and propagating 2d plasmons~\cite{Woessner_phase_shifter,Ni_limits_plasmonics}, this setup has numerous technological applications. First, it is the principal building block for electrical induction of light-sensitive $p-n$ junctions~\cite{Woessner_electrical,Castilla_fast_THz,Olbrich2016}. Second, it represents a 'mirror' for 2d plasmons~\cite{Sydoruk_gate_edge}, and mirrors formed at the two edges of the gate enable the formation of plasmonic resonant cavities~\cite{Bylinkin_tight_binding,Knap_crystals,Muravev_grating}. 

Our analysis is based on the Wiener-Hopf technique for the solution of integro-differential equations in semi-bounded domains. It has been applied since mid XX-th century to the electromagnetic scattering by metallic edges~\cite{Senior1952}, waveguide terminations~\cite{Nussenzveig_WG_termination}, and related partially bounded obstacles~\cite{Kay_reactance_discontinuity}. In late 80's, the method enabled the studies of magnetoplasmons propagating at 2DES edges~\cite{volkov1988edge}. Recently it was applied to diffraction and plasmon launching by 2DES terminations~\cite{Margetis2017,Zhang_Wiener_Hopf} and their contacts with metals~\cite{Nikulin2021}. The case of partially gated 2DES is much more complex, as the gate and 2DES have different conductivity and lie in different planes at the vertical scale. To handle this complexity, we 'absorb' the screening currents in 2DES into the re-definition of electromagnetic propagator, and reduce the problem to the scattering by a metallic wedge. A similar trick was used in the theory of plasmons in grating-gated systems~\cite{Matov1993,Mikhailov1998} and in the theory of 2D plasmons propagating along strip-like gates~\cite{Zabolotnykh_proximity}. Compared to the well-known wedge diffraction in free space~\cite{Senior1952}, the structure of fields in our case is considerably richer. The scattered fields include 2d plasmons with different wavelengths (gated and ungated) and different amplitudes running away from the gate edge. The amplitudes of plasmons are explicitly found from our analytical solution. Both amplitudes are maximized as the gate approaches the 2DES, and can readily exceed the amplitude of incident electric field. The amplitudes also possess a non-trivial maximum as a function of 2d inductance, i.e. imaginary part of 2d conductivity.

\section{Solution of the scattering problem}
The geometry of the problem is shown in Fig.~\ref{structure}. Electromagnetic wave with complex electric field amplitude ${\bf E} = {\bf E}_0 e^{ik_x x + i k_z z}$ is incident on a semi-infinite metal gate at a distance $z_0$ from the 2DES located in the plane $z=0$. The incident wave is $p$-polarized, i.e. its magnetic field ${\bf B}_0$ is parallel to the gate edge, while the electric field ${\bf E}_0$ lies in the plane of incidence. The time dependence of the field is assumed harmonic, ${\bf E}(t) = {\bf E}_0 e^{-i\omega t} + {\rm h.c.}$, and the harmonic exponent will be hitherto skipped. The metal is considered as a perfect conductor (PC), while the 2DES is characterized by a complex sheet conductivity $\sigma_{2d}$.

The scaling analysis suggests that the total field would depend in non-trivial fashion only on three dimensionless parameters: the angle of incidence $\alpha = {\rm arctan}(k_x/k_z)$, the gate-2DES separation timed by the wave number $k_0z_0 \equiv 2\pi z_0/\lambda_0$, and the 2d conductivity  \textcolor{black}{$\sigma_{2d}$  timed by the free-space impedance $\eta = \sigma_{2d}Z_0/2$ ($Z_0=377$ Ohm in SI units and $c/4\pi$ in Gaussian units)}. The total field would scale linearly with $E_0$, while the coordinates can be always scaled by the wavelength. 

To obtain the distribution of the electric field in the structure under study, we start from the wave equation for the total vector potential $\textbf{A}(x,z)$ considering the currents in 2DES ${{\mathbf{J}}_{2D}}(x)$ and metal gate ${{\mathbf{J}}_\text{g}}(x)$ as known sources. After the Fourier transform with respect to the $x$-coordinate (${{\mathbf{A}}_{\text{q}}}(z)=\int_{-\infty }^{\infty }{\mathbf{A}(x,z){{e}^{-iqx}}dx}$), the wave equation acquires the following form  
\begin{equation}
\label{eq-wave_eq}
\chi(q)^2{{\mathbf{A}}_{\text{q}}}(z)-\frac{{{\partial }^{2}}{{\mathbf{A}}_{\text{q}}}(z)}{\partial {{z}^{2}}}=2Z_0[{{\mathbf{J}}_{2\text{D, q}}}\delta (z)+{{\mathbf{J}}_{\text{g, q}}}\delta (z-{{z}_{0}})],
\end{equation}
where ${{\mathbf{J}}_{\text{g, q}}}$ and ${{\mathbf{J}}_{\text{2D, q}}}$ are the Fourier harmonics of the current density, $\chi(q)=\sqrt{q^{2}-k_{0}^{2}}$ is the decay constant of the field in the vertical direction, $c$ is the speed of light, $k_{0}=\omega/c$ is the wave number of the incident field. Each current source in \eqref{eq-wave_eq} produces an exponentially decaying field. Together with the incident wave, they form the total vector potential
\begin{multline}
\label{solution_of_wave_eq}
{{\mathbf{A}}_{\text{q}}}(z)=G(q)[{{\mathbf{J}}_{2\text{D,}\ \text{q}}}{{e}^{-\chi (q)|z|}}-{{\mathbf{J}}_{\text{g,}\ \text{q}}}{{e}^{- \chi (q)|z-{{z}_{0}}|}}]+
\\
+{{\mathbf{A}}_{\text{ext,}\ \text{q}}}e^{-i k_z z},
\end{multline}
where ${\mathbf{A}}_\text{ext, q}$ is the Fourier transform of the incident vector potential, $G(q)=Z_0/\chi (q)$ is the electromagnetic kernel in the Fourier domain. 

We proceed to eliminate the yet unknown current density in 2DES from \eqref{solution_of_wave_eq}. To this end, we express the electric field via the vector potential $\mathbf{E_q}(z)=-i {\bf q} \varphi_{\bf q}(z) + i k_0 {\bf A}_{\bf q}(z)$, link the scalar and vector potentials in the Lorentz gauge, and apply the Ohm law in 2DES  ${{\mathbf{J}}_{2\text{D,}\text{q}}}=\sigma {{\mathbf{E}}_{\text{q}}}(z=0)$. The resulting relation reads as
\begin{equation}
\label{Ohm}
    {{{J}}_{2\text{D, q}}}={\frac{\chi^{2} (q)\sigma_{2d}}{i{{k}_{0}}\varepsilon(q) }} ({{{A}}_{\text{ext,}\ \text{q}}}+G(q){{{J}}_{\text{g, q}}}{{e}^{-\chi (q){{z}_{0}}}})
\end{equation}
where $\varepsilon (q)=1+i\eta \chi (q)/{k}_{0}$ is the dielectric function of ungated 2DES. At this stage it's worth noting that 2d current density can be selectively enhanced provided $\varepsilon(q) =0$ for some value of the wave vector $q$. The effect is nothing but plasmon-mediated field enhancement.

Introducing \eqref{Ohm} into the initial wave equation \eqref{solution_of_wave_eq}, we arrive at the relation between electric field $E_{\rm q}(z)$ and current density in the gate $J_{\text{g, q}}$:
\begin{multline}
\label{eq-solve_of_wave_eq}
{E}_{\text{q}}(z)=E_\text{ext,q}\frac{{{\varepsilon }_{g}(q,z)}}{\varepsilon (q)}{{e}^{-\chi (q)|z|}}-\\
 i{{J}}_\text{g,q} Z_0 \frac{\chi(q) }{k_{0}}\frac{{\varepsilon }_{g}(q,z)}{\varepsilon (q)}{e}^{-\chi (q)|z-{{z}_{0}}|},
\end{multline}
where we have introduced the Fourier transform of the incident field
\begin{equation}
{E}_{\text{ext,q}}={{{E}}_{0x}}\left( i /\mu_{-}(q)-i /\mu_{+}(q) \right),  
\end{equation}
$E_{0x} = E_0 \cos \alpha$ is projection of the incident field vector on the gate plane, ${{\varepsilon }_{g}}(q,z)=1+i\eta \chi (q)(1-{{e}^{-2\chi (q){{z}}}})/k_0$ has the meaning of the dielectric function of gated 2DES, with the gate located at the height $z$, and $\mu _\pm(q)=k_x-q\mp i\delta$. 

\begin{figure}[ht]
\center{\includegraphics[width=0.9\linewidth]{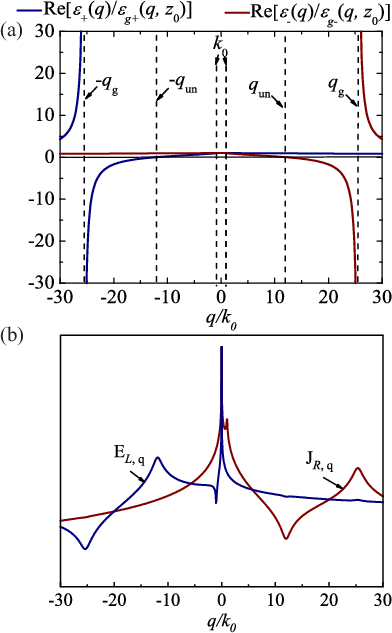}}
\caption{ Scattering problem in the spectral domain (a) ratio of factorized dielectric functions $\text{Re}[\varepsilon_{\pm}(q)/\varepsilon_{\text{g}\pm}(q)]$ as a function of wave vector, with special points at the wave vector of gated ($q_g$) and ungated ($q_u$) plasmons (b) Fourier spectra of the electric field $E_{\text{q}}(z=z_0)$ in theplane of the gate and the current density in the gate $J_{ \text{g,q}}(z=z_0)$. Numerical values are $z_0 k_0$=0.012, $\eta^{'}\ll\eta^{''}$, $\eta''=0.1$ }
\label{El_Jr}
\end{figure}

The particular scattering equation amenable for Wiener-Hopf solution is obtained from \eqref{eq-solve_of_wave_eq} by setting $z=z_0$, i.e. in the plane of gate location. We also make use of the narrow spatial spectrum of the incident field $E_\text{ext,q} \propto \delta(q-k_x)$, which allows us to set $q\approx k_x$ in the screening functions for incident field. After that, the scattering equation takes the form
\begin{equation}
\label{eq-wh-equation}
{E}_{\text{q}}(z_0)=E_\text{ext,q}\frac{{{\varepsilon }_{g}(k_x)}}{\varepsilon (k_x)}{{e}^{-ik_z z_0}}-
 i{{J}}_\text{g,q} Z_0 \frac{\chi(q) }{k_{0}}\frac{{\varepsilon }_{g}(q)}{\varepsilon (q)},
\end{equation}
where we have used the shorthand notation $\varepsilon_{g}(q,z_0) \equiv \varepsilon_{g}(q)$. The particular choice of height $z=z_0$ is dictated by the fact that electric field is bounded here to the left half-space, while the current in the gate is bounded to the right half-space. The functions ${E}_{\text{q}}(z_0)$ and ${{J}}_\text{g,q}$ have therefore easily recognizable analytic properties. Namely,  ${E}_{\text{q}}(z_0)$ is analytic in the upper half-plane of complex $q$-variable, and ${{J}}_\text{g,q}$ is analytic in the lower half-plane of complex $q$-variable.

The principal step of the Wiener-Hopf solution lies in the multiplicative decomposition of the kernel functions $\chi(q)$, $\varepsilon(q)$ and $\varepsilon_g(q)$ into the parts analytic in the lower ($-$) and upper ($+$) complex half-planes. Assuming that the surrounding medium has weak residual dissipation, $k_0 = k_0'+ik_0''$, we immediately arrive to
\begin{equation}
    \chi(q) = \chi_+(q)\chi_-(q), \qquad \chi_{\pm}(q) = \sqrt{q\pm k_0},
\end{equation}
where the branch cuts of the square root function start at $\pm k_0$ and run to $\pm i\infty$ without crossing the real axis. The decomposition of dielectric functions $\varepsilon(q)$ and $\varepsilon_g(q)$ is achieved with the Cauchy integral theorem applied to a narrow rectangle enclosing the real axis:
\begin{equation}
\label{eq-factorisation}
{{\vartheta }_{\pm }}(q)=\exp \left\{ \pm \frac{1}{2i\pi }\int\limits_{-\infty }^{\infty }{\frac{\text{Ln}\vartheta (u)du}{u-q\pm i0}} \right\},
\end{equation}
where $\vartheta(q) = \{\varepsilon(q),\varepsilon_g(q)\}$ is the function subjected to factorization.

After the factorized representations of functions in the scattering equation \eqref{eq-wh-equation} are found, one collects the terms analytic in the upper and lower complex-$q$ planes in the left and right-hand sides of the scattering equation. This results in 
\begin{gather}
\label{eq-wh-collected}
{{K}_{+}}(q){E}_{\rm q}(z_0)+{{C}_{+}}(q)={-{C}_{-}}(q)-{{K}_{-}}(q){J_{\rm g,q}},\\
{{K}_{-}}(q)=-iZ_0\frac{\chi_-(q)}{k_0}\frac{{{\varepsilon }_{g-}}(q)}{{{\varepsilon }_{-}}(q)},\\
{{K}_{+}}(q)=\frac{1}{\chi_+(q)}\frac{{{\varepsilon }_{+}}(q)}{{{\varepsilon }_{g+}}(q)},\\
 {{C}_{\pm}}(q)=\pm\frac{iE_{0x}}{\chi_+(k_x)}\frac{{{\varepsilon }_{g\pm}}(k_x)}{{{\varepsilon }_{\pm}}({{k}_{x}})}\frac{{{e}^{-ik_z z_0}}}{\mu_\pm(q)}, 
\end{gather}

Equation \eqref{eq-wh-collected} states that two functions analytic in the upper and lower half planes are equal identically in a narrow strip enclosing the real axis. It implies that they both are equal to a function $P(q)$ entire in the whole complex plane, i.e. to a polynomial. Further on, this polynomial should be exactly zero if finite values of inverse Fourier transforms $E_L(z)$ and $J_g(x)$ are demanded. This leads us to the final expression for the Fourier spectra of current in the gate and electric field in the gate plane:
\begin{gather}
\label{eq-Jr}
\frac{Z_0 J_{\rm g,q}}{E_{0x}}= \frac{-i \varepsilon_-(q){{\varepsilon }_{g-}}({{k}_{x}}){{k}_{0}}{{e}^{-ik_z z_0}}}{ {{\chi }_{+}}({{k}_{x}}){{\varepsilon }_{g-}}(q){{\varepsilon }_{-}}({{k}_{x}}){{\chi }_{-}}(q)\mu_{-}(q) },\\
\label{eq-El}
\frac{E_{\rm q}(z_0)}{E_{0x}}=\frac{-i{{\chi }_{+}}(q){{\varepsilon }_{g+}}(q){{\varepsilon }_{g-}}({{k}_{x}})e^{-ik_z z_0}}{{{\chi }_{+}}({{k}_{x}}){{\varepsilon }_{+}}(q){{\varepsilon }_{-}}({{k}_{x}})\mu_{+}(q)}
\end{gather}

The total field in the 2DES is finally obtained by introducing the auxiliary gate current, eq.~(\ref{eq-Jr}), into the fundamental solution of the wave equation, \eqref{eq-solve_of_wave_eq}: 

\begin{multline}
\label{eq-Eq_2d}
E_{\rm{q}}(z=0)=\frac{E_{\rm ext, q}}{\varepsilon ({{k}_{x}})}-\\
\frac{iE_{0x}}{\mu_-(q) \varepsilon ({{k}_{x}})} \frac{{{\chi }_{+}}(q)}{{{\chi }_{+}}({{k}_{x}})}\frac{ {{\varepsilon }_{g-}}({{k}_{x}})}{\varepsilon_+ (q){{\varepsilon }_{g-}}(q)}{{e}^{-\chi(q)z_0}}
\end{multline}

Equation \eqref{eq-Eq_2d} is the central result of our diffraction theory. Its first term is nothing but the incident field screened by the unbounded 2DES, while the second one is the field generated by semi-infinite gate. Despite relatively concise structure, \eqref{eq-Eq_2d} contains all necessary physical effects which we proceed to discuss.

\section{Structure of the diffracted fields}

\subsection{Spectral structure}

The first noteworthy property of the diffracted field in 2DES is relatively slow decay of its Fourier components as $q\rightarrow \infty$, namely $E_{\rm q}(z=0) \propto |q/k_0|^{-1/2} e^{-|q| z_0}$. The slow decay translates into large real-space field at $x=0$, $E(x=0,z=0) \sim E_{0x} \min \{(k_0 d)^{-1/2}, |\eta|^{-1/2}\}$. Such field enhancement is reminiscence of lightning-rod effect at the sharp-angled metal wedge~\cite{Landau_Electrodynamics}. Naturally, finite 2DES conductivity and  finite gate-2DES separation smear out this field singularity.

\begin{figure}[ht!]
\center{\includegraphics[width=1\linewidth]{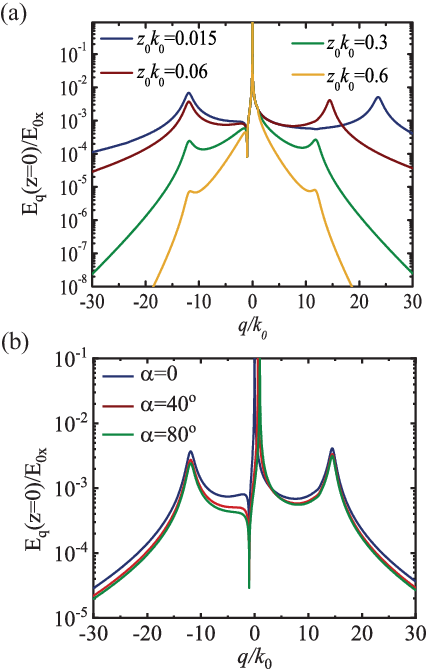}}
\caption{Spatial Fourier spectrum of electric field in the 2DES plane $E_q(z=0)$ (a) for different gate-channel separations $k_0z_0$ and fixed incidence angle $\alpha=0$ (b) for different incidence angles $\alpha$ at fixed $z_0 k_0=0.06$. In both panels, $\eta = 0.1i+0.005$}
\label{Eq_vs_q}
\end{figure}

Even more remarkable is the structure of poles in Fourier spectra of scattered fields. When transformed to real space, these poles produce running wave contributions to the electric field. As a first example, we examine the spectral structure  of $J_{\rm g,q}$ and $E_{\rm q}(z_0)$, plotted in Fig.~\ref{El_Jr} (b). The gate current has a pole at a large positive value of $q=q_g$, while the electric field in the gate plane has a pole at a negative value $q=-q_u$, $q_u<q_g$. A more detailed analysis ascribes these peak wave vectors to the propagating gated (g) and ungated (u) plasmons. These wave vectors appear as (complex) zeros of $\varepsilon_g(q)$ and $\varepsilon(q)$, respectively (see Appendix~\ref{App_plasm_basics}). In the physically relevant limit of weakly-dissipative 2DES with moderate carrier density $\eta' \ll \eta'' \ll 1$ and proximity gates $k_0z_0\ll 1$, they are approximated as $q_u \approx k_0/\eta''$, $q_g \approx k_0 / (2\eta '' k_0 z_0)^{1/2}$.

To show the correspondence between spectral peaks and plasmons, we note the presence of the screening function $\varepsilon_{g-}(q)$ in the denominator of gate current spectrum $J_{\rm g,q}$, Eq.~\eqref{eq-Jr}. According to the definition of factorized functions, $\varepsilon_{g-}(q) = \varepsilon_{g}(q)/\varepsilon_{g+}(q)$. It implies that zeros of $\varepsilon_{g-}(q)$ in the upper half-plane coincide with zeros of an ordinary screening function $\varepsilon_{g}(q)$, i.e. extended plasmons. The principal difference between partial screening function $\varepsilon_{g-}(q)$ and ordinary screening function $\varepsilon_{g}(q)$ is the single-sided positioning of zeros of the former. This fact is clarified once again in Fig.~\ref{El_Jr} (a), where we plot the factorized fractions $\varepsilon_{\pm}(q)/\varepsilon_{g\pm}(q)$. In accordance with physical intuition, the plasma oscillations in the gate run from its edge rightwards. 

A similar line of arguments can be provided to show that $E_{\rm q}(z=z_0)$ possesses a peak at the wave vector of ungated plasmons $q=-q_u$, which is the zero of $\varepsilon(q)$ with negative real part.

The peaks in the spectrally-resolved electric field in 2DES, $E_{\rm q}(z=0)$, are now double-sided. It implies that launching of 2D plasmons by the gate edge is bidirectional. This fact is illustrated in Fig.~\ref{Eq_vs_q}, where the spectra are presented for different gate-channel separations. {\textcolor{black}{The curves are highly asymmetric with respect to the positive and negative $q$, which is the consequences of different screening properties of gated part of 2DES ($q>0$) and its ungated part ($q<0$). }}

The peak wave vector for $q<0$ is insensitive to the gate position, as it should be for the ungated plasmon. The peak wave vector at $q>0$, on the contrary, is maximized at small $z_0$, in agreement with the expectation for gated plasmons. At large gate-channel separations, the Fourier spectrum $E_{\rm q}(z=0)$ becomes nearly symmetric (orange line in Fig.~\ref{Eq_vs_q}). This reflects the lack of gate screening on the properties of 2d plasmons, and indistinguishable propagation of waves in the gated and ungated sections. 

The angular dependence of diffracted fields is mainly governed by direct proportionality to $E_{0x}=E_0\cos\alpha$. The field spectra normalized by $E_{0x}$ are shown in Fig.~\ref{Eq_vs_q} (b) and do not differ considerably. The remaining angular dependence of scattered fields $E_{\rm q}-E_{\rm ext,q}$ is mainly governed by the factor $\chi_+^{-1}(k_x) = k_0^{-1/2}(1+\sin\alpha)^{-1/2}$. The other terms produce weak angular dependence, at least in the limit $|\eta|\ll 1$.

\subsection{Real-space structure and plasmon amplitudes}

Analytical insights into the amplitudes of gated and ungated plasmons launched by terminated gate can be obtained by inverse Fourier transform of the field spectrum \eqref{eq-Eq_2d}
\begin{equation}
\label{eq-inversee-Fourier}
    {{E}}(x,z=0)=\int\limits_{-\infty }^{\infty}{E_\text{q} (z=0){{e}^{iqx}}\frac{dq}{2\pi}}.
\end{equation}
When analyzing the field under the gate (i.e. at $x>0$), we close the integration path in \eqref{eq-inversee-Fourier} in the upper half-plane, where the Fourier exponential $e^{iqx}$ decays rapidly. The resulting contour integral is determined by the sum of $E_\text{q}$-residues in the UHP, which are located at $q=+q_g$ and $q=k_x$. The residue at $q=k_x$ appears exactly zero. This is natural, as the right part of the 2DES is shadowed from the incident field by a perfect conductor. The residue at $q=+q_g$ governs the amplitude of launched gated plasmons. A similar argument applied to $x<0$ shows that residue of $E_\text{q}$ at $q=-q_u$ governs the amplitude of ungated plasmons.

Before proceeding to plasmon amplitudes, we note that closure of integration loop in the UHP should be done with careful bypass of branch cut starting at $q=+k_0$ and running to $+i\infty$. The branch cut contribution is responsible for the evanescent (non-propagating) field localized near the edge. We have numerically verified that it is small compared to plasmonic fields, at least for weakly dissipative 2d systems with $\eta' \ll \eta''$. The smallness apparently does not take place for highly dissipative 2DES with  $\eta' \gg \eta''$ or capacitive 2DES with $\eta'' <0$. In both latter cases, 2D plasmonic contributions are absent at all. 

\begin{figure}[ht]
\center{\includegraphics[width=0.95\linewidth]{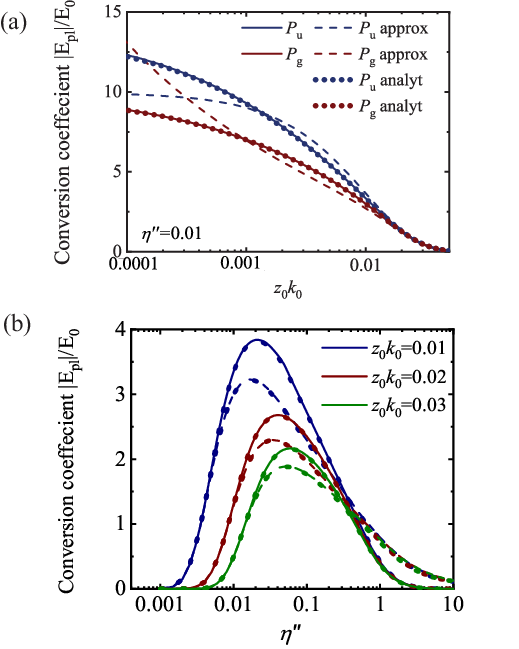}}
\caption{Amplitudes of plasmons launched by the gate edge. (a) Amplitude conversion coefficients for ungated ($P_u$) and gated ($P_\text{g}$) plasmons vs $k_0z_0$ at fixed conductivity $\eta =0.01i$. Dashed curves show the weak coupling approximation to the conversion efficiency, Eq.~\eqref{eq-conversions-app} (b) The conversion coefficient for gated (dashed curves) and ungated (solid curves) plasmons vs 2d conductivity $\eta^{''}$ for different values of $k_0z_0$. \textcolor{black}{Dotted curves at (a) and (b) show the fully analytical approximation of the conversion efficiency using absolute values of the factorized functions ~\eqref{eq_fact_analyt},~\eqref{eq_fact_analyt_g}}.
For all figures $\eta^{'}\ll\eta^{''}$.}
\label{Epl_conversion}
\end{figure}

The contour integration technique outlined above allows us to present the field in 2DES plane as
\begin{multline}
\label{eq-2des-field-spatial}
    E(x,z=0) = \frac{E_{0x} e^{ik_xx}}{\varepsilon(k_x)}\theta(-x) + E_{\rm g} e^{i q_g x} \theta(x) +\\ E_{\rm u} e^{-iq_u x} \theta(-x) + E_{\rm ev}(x),
\end{multline}
where the first term is the incident plane wave screened by 2DES in its geometrically illuminated part, $E_{\rm ev}(x)$ is the evanescent field localized near the edge, and the gated/ungated plasmon amplitudes $E_{g/u}$ are given by:
\begin{equation}
\label{eq-field-ungated}
    E_u = \left. \frac{\chi_+(q)}{\chi_+(k_x)}\frac{\varepsilon_-(q)\varepsilon_{g-}(k_x)}{\varepsilon_-(k_x)\varepsilon_{g-}(q)}\frac{E_{0x}e^{-\chi(q)z_0}}{(q-k_x)\partial\varepsilon/\partial q}\right|_{q=-q_u},
\end{equation}
\begin{equation}
\label{eq-field-gated}
    E_g = \left. \frac{\chi_+(q)}{\chi_+(k_x)}\frac{\varepsilon_{g+}(q)\varepsilon_{g-}(k_x)}{\varepsilon_+(q)\varepsilon_{-}(k_x)}\frac{E_{0x}e^{-\chi(q)z_0}}{(q-k_x)\partial\varepsilon_g/\partial q}\right|_{q=q_g}
\end{equation}
The conversion efficiencies of the incident field into gated and ungated plasmons (i.e. the ratios $P_{g/u}=E_{g/u}/E_0$) computed with Eqs.~\eqref{eq-field-ungated} and \eqref{eq-field-gated} are shown in Fig.~\ref{Epl_conversion}. We stick to the case of the normal incidence $\alpha=0$ due to its ubuquitios presence in the experiment. The calculations show that both conversion coefficients grow as the gate is proximized with 2DES [Fig.~\ref{Epl_conversion} (a)]. The amplitude of \textcolor{black}{both} plasmons saturates for close proximity at $z_0 k_0 \ll \eta''$. It is possible to show that the conversion coefficient of ungated plasmons at $k_0z_0 = 1$ equals to that for direct metal-2DES contact~\cite{Nikulin2021}.

\begin{figure}[ht]
\center{\includegraphics[width=0.95\linewidth]{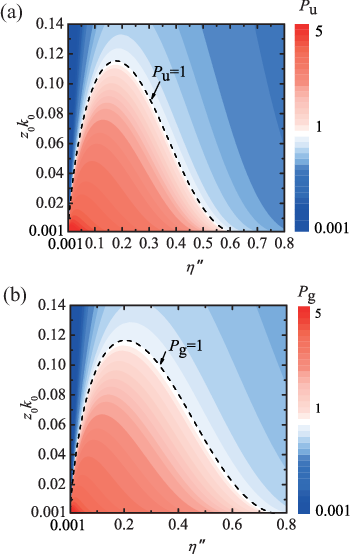}}
\caption{Color map of the conversion efficiency as a function of gate-channel separation $k_0z_0$ and 2d conductivity $\eta^{''}$ for (a) ungated ($P_u$) and (b) gated ($P_g$) plasmons. For all figures $\eta^{'}\ll\eta^{''}$. Dashed black line marks the unity conversion efficiency}
\label{Epl_conversion_3D}
\end{figure}

The dependences of conversion coefficients on 2d conductivity $\eta''$ appear highly non-trivial, as displayed in Fig.~\ref{Epl_conversion} (b). Both ungated (solid) and gated (dashed) amplitudes grow for small surface conductivity, reach a maximum for some 'optimum conductivity' $\eta''_{\rm opt}$, and drop down at very large $\eta''$ due to the mirroring of the incident field by highly conductive 2DES.
The 'optimum' conductivity $\eta''_{\rm opt}$ scales linearly with $k_0z_0$ for both gated and ungated waves, though the numerical values of optima for the two waves are slightly different. Interestingly, the conversion coefficient into the ungated plasmons exceeds that for gated ones at small $\eta''$, while for $\eta''\gg 1$ the situation is reversed. The full dependences of conversion coefficients $P_{g/u}$ on both $\eta''$ and $k_0z_0$ are shown in Fig.~\ref{Epl_conversion_3D}.

While the results for plasmon amplitudes (\ref{eq-field-ungated}, \ref{eq-field-gated}) are fully analytical, the evaluation of emerging factorized functions $\varepsilon_{\pm}(q)$ may be difficult. We further provide even more transparent expressions for the wave amplitudes in the practically interesting case of ultra-confined plasmons $\eta'' \ll 1$. In this limit, both dielectric functions $\varepsilon(q)$, $\varepsilon_g(q)$ and their factorized components $\varepsilon_{\pm}(q)$, $\varepsilon_{g\pm}(q)$ are close to unity. Setting them to unity identically, we arrive at approximate conversion coefficients
\begin{equation}
\label{eq-conversions-app}
    {{E}^{\rm app}_{\rm g/u}}=-\left.\frac{{e}^{-\chi \left( q \right)z_0}}{{{ \partial \varepsilon_{\rm g/u} / \partial q }}}\frac{{E}_{0x} }{q- {k}_{x}}\frac{\chi_+(q)}{\chi_+(k_x)}\right|_{q = \pm q_{\rm g/u}}.
\end{equation}

The same result can be obtained bypassing the complex Wiener-Hopf technique, and assuming the current distribution in the gate ${\bf J}_g(x)$ to be unaffected by 2DES (see Appendix B for detailed derivation). The particular from of ${\bf J}_g(x)$ can be borrowed from the theory of diffraction at a semi-infinite metallic sheet~\cite{Noble1958MethodsBO}. The total non-uniform field created by the incident wave and semi-infinite gate is then screened by 2DES, which amounts to its multiplication by $\varepsilon^{-1}(q,\omega)$ in Fourier domain. Extraction of plasmon pole amplitude from the total field leads exactly to the expressions (\ref{eq-conversions-app}).

The approximate conversion coefficients (\ref{eq-conversions-app}) are shown in Fig.~\ref{Epl_conversion} with dashed lines. They agree well with complete expressions at relatively large gate-channel separations $k_0z_0 \gtrsim \eta''$. Agreement at large distances (or small surface conductivity) is well anticipated, as the 2DES in this case makes no back-action on the surface currents in the gate. 

For ultra-proximized gates $k_0z_0 \lesssim \eta \ll 1$, the approximate scheme breaks down, and more accurate approximation of $\varepsilon_{\pm}(q)$ is required. To achieve this, we expand the logarithm $\ln \varepsilon(q)$ in powers of small $\eta''$ in (\ref{eq-factorisation}), and retain only the terms contributing to the absolute values of factorized functions. This results in

\textcolor{black}{
\begin{equation}
\label{eq_fact_analyt}
    |\varepsilon_{\pm}(q)|\approx \sqrt{|\varepsilon(q)|\left|\frac{q\pm q_{u}}{q \mp q_{u}}\right|} \exp\left(\pm\frac{\eta^{''}(q-|\chi(q)|\text{sgn}(q)}{2\pi k_0}\right),
    \end{equation}
\begin{multline}
\label{eq_fact_analyt_g}
    |\varepsilon_{g\pm}(q)|\approx \sqrt{|\varepsilon_g(q)\left|\frac{q \pm q_{g}}{q \mp q_{g}}\right|}  \times \\    \exp\left(\pm\frac{2z_0\eta^{''}}{\pi}\left[|\chi(q)|\ln \left|\frac{\chi_{-}(q)}{\chi_{+}(q)}\right|+q\right]\right).
   \end{multline}}
\textcolor{black}{The approximate factorized 
 dielectric functions (\ref{eq_fact_analyt}),(\ref{eq_fact_analyt_g}) are perfectly suited for the description of plasmon conversion efficiencies in a broad range of parameters. The respective conversion efficiencies match well the full expressions at arbitrary gate-channel separation (dotted lines in Fig.~\ref{Epl_conversion}(a)) and even for $\eta'' \gtrsim 1$ (dotted lines in Fig.~\ref{Epl_conversion}(b)).} 
 
\subsection{Comparison with simulations}

The complexity of Wiener-Hopf method in electromagnetic problems calls for its verification by comparison with other techniques. One such comparison in the limit of 'weak coupling' between gate and 2DES was provided in the previous section. Here, we address the method's validity by direct simulations of electromagnetic wave scattering in the partially gated 2DES.

\begin{figure}[h!]
\center{\includegraphics[width=0.95\linewidth]{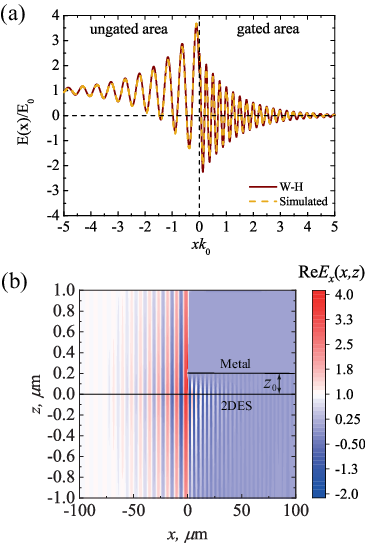}}
\caption{(a) The electric field in 2DES in dependence of coordinate x normalized to the free space wave vector $k_0$ for distances $z_0=0.2$ $\mu$m from 2DES to the gate and (b) The electric field spatial distribution in studied structure for frequency $\omega/2 \pi$=3 THz}
\label{Simulation}
\end{figure}

The simulation was performed using the finite element method realized in CST Microwave Studio package. The case of normal incidence at a frequency of 3 THz was considered. The gate-channel separation was taken as $z_0 = 0.2$ $\mu$m, the 2DES conductivity was considered as inductive with numerical value $\eta =0.004+0.08i$. The simulated electric field profile in 2DES is shown in Fig.~\ref{Simulation}(a) with dashed orange line. As expected, it displays fast oscillations associated with gated plasmons at $x>0$, and somewhat slower oscillations associated with ungated plasmons at $x<0$. The plateau in electric field at $k_0x \ll -1$ is the field of the incident wave. \textcolor{black}{The three-dimensional mapping of the field (Fig. \ref{Simulation}(b)) reveals tight confinement of the gated mode in the $z$-direction, as compared to the ungated one. This result is well anticipated as the vertical localization length, in the non-retarded limit, is the inverse of wave vector.}

The total simulated field agrees well with the Wiener-Hopf result \eqref{eq-2des-field-spatial}, where the plasma wave amplitudes $E_{g/u}$ are given by Eqs.~(\ref{eq-field-ungated},\ref{eq-field-gated}). As for the evanescent contribution $E_{ev}(x)$ to the total field, we did not attempt to evaluate it, and it was set identically zero in our comparison. The Wiener-Hopf result with $E_{ev}(x)\equiv0$ is shown in Fig.~\ref{Simulation} with the red solid line, and matches well the numerical simulation shown with orange dashed line. 

\section{Conclusion and future outlook}

We have obtained an exact solution for electromagnetic wave scattering at at partially gated uniformly doped two-dimensional electron system. The principal result of this solution is the bidirectional launching of two-dimensional plasmons by the junction for 2DES with large inductive conductivity $\sigma_{2d}''>\sigma_{2d}'$. The amplitudes of launched plasmons, both gated and ungated, were explicitly obtained. The electric field amplitude of the gated plasmon always exceeds that of the ungated. The amplitudes of both waves can exceed the amplitude of incident electric field for proximate gates $k_0z_0 \ll 1$. The 2d conductivity providing the maximization of plasmons' amplitude is given, by the order of magnitude, by $\sigma''_{2d} \sim k_0 z_0/Z_0$.

As the problem deals with harmonic single-frequency excitation (as in most spectroscopic and photovoltaic experiments), the frequency dependence of 2d conductivity $\sigma_{2d}(\omega)$ is immaterial for the present calculation. What really matters is the numerical value of $\sigma_{2d}$ at given excitation frequency. It can be obtained either from independent calculations (e.g. using the kinetic equation or Kubo formula~\cite{Hwang2007a}) or from spectroscopic~\cite{Zhukova_spectroscopy} or ellipsometric~\cite{toksumakov2023anomalous} experiments. Most remarkably, the conductivity in the current setting can be non-local or, in other words, its Fourier transform can depend on $q$ explicitly. The possibility to treat non-local effects in conductivity with the present method is guaranteed by the uniformity of 2DES doping.

The scattering on non-uniformly doped 2DES, on the contrary, requires a considerable modification of the present technique. The system of scattering equation for such 2DES involves four unknown functions: the electric fields in the left and right sections of the 2DES plane and of the gate plane. The solution of such {\it system} of the Wiener-Hopf equation is challenging, as compared to a single equation we have solved here. Such non-uniform doping of 2DES appears upon application of voltage to a semi-infinite gate~\cite{Olbrich2016}.

Despite its apparent limitations, the presented solution can be used as a basic block for modeling of optoelectronic devices based on gated 2DES. Known the electric field, one readily evaluates the absorbed electromagnetic power, electron heating and bolometric effects, and various non-linear photovoltaic phenomena. One such photovoltaic effect is the photon drag, whose magnitude is proportional to the asymmetry of electric field in the Fourier domain~\cite{Fateev_rectification} $|E(q)|^2-|E(-q)|^2$.

\textcolor{black}{Our exact solutions for the total electromagnetic field in the 2DES (\ref{eq-Eq_2d}) and the plasmon amplitudes (\ref{eq-field-ungated}), (\ref{eq-field-gated}) directly apply to nonlocal models of conduction in 2DES. In that case, the normalized 2DES conductivity $\eta$ should be treated as wave vector-dependent function $\eta(q)$. These effects are important for description of ultra-confined 2D plasmons with velocity approaching the Fermi velocity~\cite{Lundeberg2017c}. Another physical situation when non-locality matters appears in 2DES with dc current. Such structures can amplify the incident radiation~\cite{Mikhailov1998} and act as 'gain media' for 2d plasmons~\cite{moiseenko2024dissipative,Sydoruk_mirrors}, which can enable the creation of compact radiation sources. We note that non-local conductivity models are currently unavailable in commercial electromagnetic simulators, which raises the value of the exact solution obtained.}

While the present calculation dealt with free-space excitation, other types of electromagnetic sources can be considered with minor modifications. Of particular interest is the scattering of plasma wave (gated or ungated) at the gate termination. The problem gained considerable interest due to applications of grating gate plasmons for terahertz detection~\cite{Sydoruk_gate_edge,Aizin_finite,Aizin_NPhoton,Knap_crystals,Gorbenko_LateralPC}. The plasmon scattering problem in such setting was considered previously either fully numerically~\cite{Sydoruk_gate_edge}, or using approximate plane-wave matching techniques~\cite{Aizin_finite,Gorbenko_LateralPC}.

This work was supported by the Russian Science Foundation (Grant No. 24-79-00094).

\appendix
\section{The plasmon properties of ungated and gated 2DES}
\label{App_plasm_basics}
We review the properties of plasmons in structures described by dielectric functions for gated and ungated 2DES. The wave vectors of ungated and gated plasmons are determined from the solutions of equations $\varepsilon(q)=0$ and $\varepsilon_g(q,z_0)=0$, respectively. The determination of dispersion in the form $\omega(q)$ requires the specification of frequency-dependent conductivity. We choose it, in the present Appendix, in the Drude form $\sigma_{2d} = i ne^2/m(\omega+i/\tau)$.
Equation $\varepsilon(q)=0$ (ungated plasmons) is readily solved both with respect to wave vector and frequency. In the former case, the dispersion reads as $q=k_0(1-1/\eta^2)^{1/2}.$

\begin{figure}[ht]
    \centering
    \includegraphics[width=0.99\linewidth]{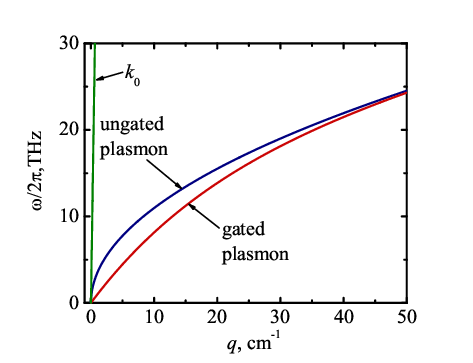}
    \caption{Dispersion of the light in the free space and dispersion of ungated and gated plasmons in 2DES with electrons concentration $n=2\times10^{12}$ $\text{cm}^{-2}$, momentum relaxation time $\tau$=1 ps and effective mass $m=0.068m_e$ (corresponding to GaAs), where $m_e$ is the electron mass.}
    \label{Dispersion_f_q}
\end{figure}

The dispersion curves for both plasmons merge at large $q z_0 \gg 1$ since the gate screening factor $e^{-2\chi(q)z_0}$ becomes negligible in that limit. At small $q/k_0$ the dispersion curves differ due to the effect of screening on the spectrum of gated plasmons (Fig.~\ref{Dispersion_f_q}), the dispersion for which of a gate close to 2DES can be described as $q_\text{g} \approx k_0 / (2\eta '' k_0 z_0)^{1/2}.$

Thus, the dielectric functions describe well the spectrum of ungated/gated plasmons, along with the well-known expressions for plasmon dispersion in ungated~\cite{Stern1967}  $\omega=\sqrt{e^2nq/2m\varepsilon}$ and gated ~\cite{Allen1977}  $\omega=\sqrt{e^2nq/m\varepsilon[1+ \text{ch}(qd)]}$  2DES, where $\varepsilon$ is the dielectric permittivity of the media surrounding 2DES.
\section{Conversion efficiency approximation for close and far located gate from the 2DES }

Here we obtain an expression for the electric field in a 2DES screened by a perfect conducting metal located far from the 2DES. In case of $z_0\gg \lambda_\text{pl}=2\pi/q_\text{pl}$ the influence of the 2DES on the current in the metal can be neglected and the plasmonic properties of such a system are described by the dielectric function $\varepsilon(q)$. In this case, the total incident field can be represented as the sum of the field of the incident wave itself and the field obtained as a result of solving the problem of wave diffraction on a metal half-plane in the absence of 2DES. Then the total electric field in 2DES can be written as:

\begin{equation}
    E\left( q \right)=\frac{{{E}_\text{ext}}\left( q \right)}{\varepsilon \left( q \right)}+\frac{i{{E}_\text{0}}{{e}^{-\chi \left( q \right)z_0}}}{\varepsilon \left( q \right){\mu_+({q}_{\text{}})}  }\frac{\chi_+ \left( q_\text{} \right)}{\chi_+ \left( k_x\right)}.
    \label{Eapprox}
\end{equation}
The approach used is also applicable to screened systems described by a dielectric function $ \varepsilon_\text{g}(q,z_0)$, since the dispersion of gated and ungated plasmons is indistinguishable at a remote gate. Calculation of the residue for (\ref{Eapprox}) at the plasmon pole $q =q _\text{g/u}$, we obtain an expression for the plasmon amplitudes
\begin{multline}
 {{E}_\text{g/u}}=\underset{q\to {{q}_{\text{g/u}}}}{\mathop{i\operatorname{Re}\text{s}}}\,E(q)=
 \\
 =\frac{-1}{{{\left. \partial \varepsilon_\text{g/u}(q) \text{/}\partial q \right|}_{q={{q}_{\text{g/u}}}}}}\frac{{{E}_{0}} {{e}^{-\chi \left( {{q}_{\text{g/u}}} \right){{z}_{0}}}}}{{({q}_{\text{g/u}}-k_x)}}\frac{\chi_+ \left(  q_\text{g/u} \right)}{\chi_+ \left( k_x\right)}.
 \label{Epl_approx}
\end{multline}
\textcolor{black}{The approximate method describes the amplitudes of both plasmons quite well in the case of a remote gate up to $k_0z_0\approx\eta^{''}$ (Fig. \ref{Epl_conversion} (a)) and allows one to avoid the complete solution of the diffraction problem in a 2DES partially screened by a metal gate using the Wiener-Hopf technique.}

\bibliography{apssamp.bib}

\providecommand{\noopsort}[1]{}\providecommand{\singleletter}[1]{#1}%
\begin{thebibliography}{50}%
\makeatletter
\providecommand \@ifxundefined [1]{%
 \@ifx{#1\undefined}
}%
\providecommand \@ifnum [1]{%
 \ifnum #1\expandafter \@firstoftwo
 \else \expandafter \@secondoftwo
 \fi
}%
\providecommand \@ifx [1]{%
 \ifx #1\expandafter \@firstoftwo
 \else \expandafter \@secondoftwo
 \fi
}%
\providecommand \natexlab [1]{#1}%
\providecommand \enquote  [1]{``#1''}%
\providecommand \bibnamefont  [1]{#1}%
\providecommand \bibfnamefont [1]{#1}%
\providecommand \citenamefont [1]{#1}%
\providecommand \href@noop [0]{\@secondoftwo}%
\providecommand \href [0]{\begingroup \@sanitize@url \@href}%
\providecommand \@href[1]{\@@startlink{#1}\@@href}%
\providecommand \@@href[1]{\endgroup#1\@@endlink}%
\providecommand \@sanitize@url [0]{\catcode `\\12\catcode `\$12\catcode `\&12\catcode `\#12\catcode `\^12\catcode `\_12\catcode `\%12\relax}%
\providecommand \@@startlink[1]{}%
\providecommand \@@endlink[0]{}%
\providecommand \url  [0]{\begingroup\@sanitize@url \@url }%
\providecommand \@url [1]{\endgroup\@href {#1}{\urlprefix }}%
\providecommand \urlprefix  [0]{URL }%
\providecommand \Eprint [0]{\href }%
\providecommand \doibase [0]{http://dx.doi.org/}%
\providecommand \selectlanguage [0]{\@gobble}%
\providecommand \bibinfo  [0]{\@secondoftwo}%
\providecommand \bibfield  [0]{\@secondoftwo}%
\providecommand \translation [1]{[#1]}%
\providecommand \BibitemOpen [0]{}%
\providecommand \bibitemStop [0]{}%
\providecommand \bibitemNoStop [0]{.\EOS\space}%
\providecommand \EOS [0]{\spacefactor3000\relax}%
\providecommand \BibitemShut  [1]{\csname bibitem#1\endcsname}%
\let\auto@bib@innerbib\@empty
\bibitem [{\citenamefont {Stern}(1967)}]{Stern1967}%
  \BibitemOpen
  \bibfield  {author} {\bibinfo {author} {\bibfnamefont {Frank}\ \bibnamefont {Stern}},\ }\bibfield  {title} {\enquote {\bibinfo {title} {{Polarizability of a Two-Dimensional Electron Gas}},}\ }\href {\doibase 10.1103/PhysRevLett.18.546} {\bibfield  {journal} {\bibinfo  {journal} {Physical Review Letters}\ }\textbf {\bibinfo {volume} {18}},\ \bibinfo {pages} {546--548} (\bibinfo {year} {1967})}\BibitemShut {NoStop}%
\bibitem [{\citenamefont {Allen}\ \emph {et~al.}(1977)\citenamefont {Allen}, \citenamefont {Tsui},\ and\ \citenamefont {Logan}}]{Allen1977}%
  \BibitemOpen
  \bibfield  {author} {\bibinfo {author} {\bibfnamefont {S.~J.}\ \bibnamefont {Allen}}, \bibinfo {author} {\bibfnamefont {D.~C.}\ \bibnamefont {Tsui}}, \ and\ \bibinfo {author} {\bibfnamefont {R.~A.}\ \bibnamefont {Logan}},\ }\bibfield  {title} {\enquote {\bibinfo {title} {{Observation of the two-dimensional plasmon in silicon inversion layers}},}\ }\href {\doibase 10.1103/PhysRevLett.38.980} {\bibfield  {journal} {\bibinfo  {journal} {Physical Review Letters}\ }\textbf {\bibinfo {volume} {38}},\ \bibinfo {pages} {980--983} (\bibinfo {year} {1977})}\BibitemShut {NoStop}%
\bibitem [{\citenamefont {Ryzhii}\ \emph {et~al.}(2007)\citenamefont {Ryzhii}, \citenamefont {Satou},\ and\ \citenamefont {Otsuji}}]{Ryzhii2007a}%
  \BibitemOpen
  \bibfield  {author} {\bibinfo {author} {\bibfnamefont {V.}~\bibnamefont {Ryzhii}}, \bibinfo {author} {\bibfnamefont {A.}~\bibnamefont {Satou}}, \ and\ \bibinfo {author} {\bibfnamefont {T.}~\bibnamefont {Otsuji}},\ }\bibfield  {title} {\enquote {\bibinfo {title} {{Plasma waves in two-dimensional electron-hole system in gated graphene heterostructures}},}\ }\href {\doibase 10.1063/1.2426904} {\bibfield  {journal} {\bibinfo  {journal} {Journal of Applied Physics}\ }\textbf {\bibinfo {volume} {101}},\ \bibinfo {pages} {024509} (\bibinfo {year} {2007})}\BibitemShut {NoStop}%
\bibitem [{\citenamefont {Woessner}\ \emph {et~al.}(2015)\citenamefont {Woessner}, \citenamefont {Lundeberg}, \citenamefont {Gao}, \citenamefont {Principi}, \citenamefont {Alonso-Gonz{\'{a}}lez}, \citenamefont {Carrega}, \citenamefont {Watanabe}, \citenamefont {Taniguchi}, \citenamefont {Vignale}, \citenamefont {Polini}, \citenamefont {Hone}, \citenamefont {Hillenbrand},\ and\ \citenamefont {Koppens}}]{Woessner2015}%
  \BibitemOpen
  \bibfield  {author} {\bibinfo {author} {\bibfnamefont {Achim}\ \bibnamefont {Woessner}}, \bibinfo {author} {\bibfnamefont {Mark~B.}\ \bibnamefont {Lundeberg}}, \bibinfo {author} {\bibfnamefont {Yuanda}\ \bibnamefont {Gao}}, \bibinfo {author} {\bibfnamefont {Alessandro}\ \bibnamefont {Principi}}, \bibinfo {author} {\bibfnamefont {Pablo}\ \bibnamefont {Alonso-Gonz{\'{a}}lez}}, \bibinfo {author} {\bibfnamefont {Matteo}\ \bibnamefont {Carrega}}, \bibinfo {author} {\bibfnamefont {Kenji}\ \bibnamefont {Watanabe}}, \bibinfo {author} {\bibfnamefont {Takashi}\ \bibnamefont {Taniguchi}}, \bibinfo {author} {\bibfnamefont {Giovanni}\ \bibnamefont {Vignale}}, \bibinfo {author} {\bibfnamefont {Marco}\ \bibnamefont {Polini}}, \bibinfo {author} {\bibfnamefont {James}\ \bibnamefont {Hone}}, \bibinfo {author} {\bibfnamefont {Rainer}\ \bibnamefont {Hillenbrand}}, \ and\ \bibinfo {author} {\bibfnamefont {Frank H.~L.}\ \bibnamefont {Koppens}},\ }\bibfield  {title} {\enquote {\bibinfo {title} {{Highly confined low-loss plasmons
  in graphene–boron nitride heterostructures}},}\ }\href {\doibase 10.1038/nmat4169} {\bibfield  {journal} {\bibinfo  {journal} {Nature Materials}\ }\textbf {\bibinfo {volume} {14}},\ \bibinfo {pages} {421--425} (\bibinfo {year} {2015})}\BibitemShut {NoStop}%
\bibitem [{\citenamefont {Chaplik}(1972)}]{chaplik1972possible}%
  \BibitemOpen
  \bibfield  {author} {\bibinfo {author} {\bibfnamefont {AV}~\bibnamefont {Chaplik}},\ }\bibfield  {title} {\enquote {\bibinfo {title} {Possible crystallization of charge carriers in low-density inversion layers},}\ }\href@noop {} {\bibfield  {journal} {\bibinfo  {journal} {Soviet Journal of Experimental and Theoretical Physics}\ }\textbf {\bibinfo {volume} {35}},\ \bibinfo {pages} {395} (\bibinfo {year} {1972})}\BibitemShut {NoStop}%
\bibitem [{\citenamefont {Nehls}\ \emph {et~al.}(1996)\citenamefont {Nehls}, \citenamefont {Schmidt}, \citenamefont {Merkt}, \citenamefont {Heitmann}, \citenamefont {Norman},\ and\ \citenamefont {Stradling}}]{Heitmann_Fermi_pressure}%
  \BibitemOpen
  \bibfield  {author} {\bibinfo {author} {\bibfnamefont {J.}~\bibnamefont {Nehls}}, \bibinfo {author} {\bibfnamefont {T.}~\bibnamefont {Schmidt}}, \bibinfo {author} {\bibfnamefont {U.}~\bibnamefont {Merkt}}, \bibinfo {author} {\bibfnamefont {D.}~\bibnamefont {Heitmann}}, \bibinfo {author} {\bibfnamefont {A.}~\bibnamefont {Norman}}, \ and\ \bibinfo {author} {\bibfnamefont {R.}~\bibnamefont {Stradling}},\ }\bibfield  {title} {\enquote {\bibinfo {title} {{Direct manifestation of the Fermi pressure in a two-dimensional electron system}},}\ }\href {\doibase 10.1103/PhysRevB.54.7651} {\bibfield  {journal} {\bibinfo  {journal} {Physical Review B}\ }\textbf {\bibinfo {volume} {54}},\ \bibinfo {pages} {7651--7653} (\bibinfo {year} {1996})}\BibitemShut {NoStop}%
\bibitem [{\citenamefont {Lundeberg}\ \emph {et~al.}(2017)\citenamefont {Lundeberg}, \citenamefont {Gao}, \citenamefont {Asgari}, \citenamefont {Tan}, \citenamefont {Duppen}, \citenamefont {Autore}, \citenamefont {Alonso-Gonz{\'{a}}lez}, \citenamefont {Woessner}, \citenamefont {Watanabe}, \citenamefont {Taniguchi}, \citenamefont {Hillenbrand}, \citenamefont {Hone}, \citenamefont {Polini},\ and\ \citenamefont {Koppens}}]{Lundeberg2017c}%
  \BibitemOpen
  \bibfield  {author} {\bibinfo {author} {\bibfnamefont {Mark~B.}\ \bibnamefont {Lundeberg}}, \bibinfo {author} {\bibfnamefont {Yuanda}\ \bibnamefont {Gao}}, \bibinfo {author} {\bibfnamefont {Reza}\ \bibnamefont {Asgari}}, \bibinfo {author} {\bibfnamefont {Cheng}\ \bibnamefont {Tan}}, \bibinfo {author} {\bibfnamefont {Ben~Van}\ \bibnamefont {Duppen}}, \bibinfo {author} {\bibfnamefont {Marta}\ \bibnamefont {Autore}}, \bibinfo {author} {\bibfnamefont {Pablo}\ \bibnamefont {Alonso-Gonz{\'{a}}lez}}, \bibinfo {author} {\bibfnamefont {Achim}\ \bibnamefont {Woessner}}, \bibinfo {author} {\bibfnamefont {Kenji}\ \bibnamefont {Watanabe}}, \bibinfo {author} {\bibfnamefont {Takashi}\ \bibnamefont {Taniguchi}}, \bibinfo {author} {\bibfnamefont {Rainer}\ \bibnamefont {Hillenbrand}}, \bibinfo {author} {\bibfnamefont {James}\ \bibnamefont {Hone}}, \bibinfo {author} {\bibfnamefont {Marco}\ \bibnamefont {Polini}}, \ and\ \bibinfo {author} {\bibfnamefont {Frank H.L.~L}\ \bibnamefont {Koppens}},\ }\bibfield  {title} {\enquote
  {\bibinfo {title} {{Tuning quantum nonlocal effects in graphene plasmonics}},}\ }\href {\doibase 10.1126/science.aan2735} {\bibfield  {journal} {\bibinfo  {journal} {Science}\ }\textbf {\bibinfo {volume} {357}},\ \bibinfo {pages} {eaan2735} (\bibinfo {year} {2017})}\BibitemShut {NoStop}%
\bibitem [{\citenamefont {Muravev}\ and\ \citenamefont {Kukushkin}(2012)}]{Muravev_detector}%
  \BibitemOpen
  \bibfield  {author} {\bibinfo {author} {\bibfnamefont {V.~M.}\ \bibnamefont {Muravev}}\ and\ \bibinfo {author} {\bibfnamefont {I.~V.}\ \bibnamefont {Kukushkin}},\ }\bibfield  {title} {\enquote {\bibinfo {title} {{Plasmonic detector/spectrometer of subterahertz radiation based on two-dimensional electron system with embedded defect}},}\ }\href {\doibase 10.1063/1.3688049} {\bibfield  {journal} {\bibinfo  {journal} {Applied Physics Letters}\ }\textbf {\bibinfo {volume} {100}},\ \bibinfo {pages} {082102} (\bibinfo {year} {2012})}\BibitemShut {NoStop}%
\bibitem [{\citenamefont {Woessner}\ \emph {et~al.}(2017{\natexlab{a}})\citenamefont {Woessner}, \citenamefont {Parret}, \citenamefont {Davydovskaya}, \citenamefont {Gao}, \citenamefont {Wu}, \citenamefont {Lundeberg}, \citenamefont {Nanot}, \citenamefont {Alonso-Gonz{\'{a}}lez}, \citenamefont {Watanabe}, \citenamefont {Taniguchi}, \citenamefont {Hillenbrand}, \citenamefont {Fogler}, \citenamefont {Hone},\ and\ \citenamefont {Koppens}}]{Woessner_electrical}%
  \BibitemOpen
  \bibfield  {author} {\bibinfo {author} {\bibfnamefont {Achim}\ \bibnamefont {Woessner}}, \bibinfo {author} {\bibfnamefont {Romain}\ \bibnamefont {Parret}}, \bibinfo {author} {\bibfnamefont {Diana}\ \bibnamefont {Davydovskaya}}, \bibinfo {author} {\bibfnamefont {Yuanda}\ \bibnamefont {Gao}}, \bibinfo {author} {\bibfnamefont {Jhih-Sheng}\ \bibnamefont {Wu}}, \bibinfo {author} {\bibfnamefont {Mark~B.}\ \bibnamefont {Lundeberg}}, \bibinfo {author} {\bibfnamefont {S{\'{e}}bastien}\ \bibnamefont {Nanot}}, \bibinfo {author} {\bibfnamefont {Pablo}\ \bibnamefont {Alonso-Gonz{\'{a}}lez}}, \bibinfo {author} {\bibfnamefont {Kenji}\ \bibnamefont {Watanabe}}, \bibinfo {author} {\bibfnamefont {Takashi}\ \bibnamefont {Taniguchi}}, \bibinfo {author} {\bibfnamefont {Rainer}\ \bibnamefont {Hillenbrand}}, \bibinfo {author} {\bibfnamefont {Michael~M.}\ \bibnamefont {Fogler}}, \bibinfo {author} {\bibfnamefont {James}\ \bibnamefont {Hone}}, \ and\ \bibinfo {author} {\bibfnamefont {Frank H.~L.}\ \bibnamefont {Koppens}},\
  }\bibfield  {title} {\enquote {\bibinfo {title} {{Electrical detection of hyperbolic phonon-polaritons in heterostructures of graphene and boron nitride}},}\ }\href {\doibase 10.1038/s41699-017-0031-5} {\bibfield  {journal} {\bibinfo  {journal} {npj 2D Materials and Applications}\ }\textbf {\bibinfo {volume} {1}},\ \bibinfo {pages} {25} (\bibinfo {year} {2017}{\natexlab{a}})}\BibitemShut {NoStop}%
\bibitem [{\citenamefont {Castilla}\ \emph {et~al.}(2024)\citenamefont {Castilla}, \citenamefont {Agarwal}, \citenamefont {Vangelidis}, \citenamefont {Bludov}, \citenamefont {Iranzo}, \citenamefont {Grabulosa}, \citenamefont {Ceccanti}, \citenamefont {Vasilevskiy}, \citenamefont {Kumar}, \citenamefont {Janzen}, \citenamefont {Edgar}, \citenamefont {Watanabe}, \citenamefont {Taniguchi}, \citenamefont {Peres}, \citenamefont {Lidorikis},\ and\ \citenamefont {Koppens}}]{Castilla_electrical_spectroscopy}%
  \BibitemOpen
  \bibfield  {author} {\bibinfo {author} {\bibfnamefont {Sebasti{\'{a}}n}\ \bibnamefont {Castilla}}, \bibinfo {author} {\bibfnamefont {Hitesh}\ \bibnamefont {Agarwal}}, \bibinfo {author} {\bibfnamefont {Ioannis}\ \bibnamefont {Vangelidis}}, \bibinfo {author} {\bibfnamefont {Yuliy}\ \bibnamefont {Bludov}}, \bibinfo {author} {\bibfnamefont {David~Alcaraz}\ \bibnamefont {Iranzo}}, \bibinfo {author} {\bibfnamefont {Adri{\`{a}}}\ \bibnamefont {Grabulosa}}, \bibinfo {author} {\bibfnamefont {Matteo}\ \bibnamefont {Ceccanti}}, \bibinfo {author} {\bibfnamefont {Mikhail~I.}\ \bibnamefont {Vasilevskiy}}, \bibinfo {author} {\bibfnamefont {Roshan~Krishna}\ \bibnamefont {Kumar}}, \bibinfo {author} {\bibfnamefont {Eli}\ \bibnamefont {Janzen}}, \bibinfo {author} {\bibfnamefont {James~H.}\ \bibnamefont {Edgar}}, \bibinfo {author} {\bibfnamefont {Kenji}\ \bibnamefont {Watanabe}}, \bibinfo {author} {\bibfnamefont {Takashi}\ \bibnamefont {Taniguchi}}, \bibinfo {author} {\bibfnamefont {Nuno M.~R.}\ \bibnamefont {Peres}}, \bibinfo
  {author} {\bibfnamefont {Elefterios}\ \bibnamefont {Lidorikis}}, \ and\ \bibinfo {author} {\bibfnamefont {Frank H.~L.}\ \bibnamefont {Koppens}},\ }\bibfield  {title} {\enquote {\bibinfo {title} {{Electrical Spectroscopy of Polaritonic Nanoresonators}},}\ }\href {\doibase 10.1038/s41467-024-52838-w} {\bibfield  {journal} {\bibinfo  {journal} {Nature Communications}\ ,\ \bibinfo {pages} {1--8}} (\bibinfo {year} {2024})}\BibitemShut {NoStop}%
\bibitem [{\citenamefont {Bandurin}\ \emph {et~al.}(2018)\citenamefont {Bandurin}, \citenamefont {Svintsov}, \citenamefont {Gayduchenko}, \citenamefont {Xu}, \citenamefont {Principi}, \citenamefont {Moskotin}, \citenamefont {Tretyakov}, \citenamefont {Yagodkin}, \citenamefont {Zhukov}, \citenamefont {Taniguchi}, \citenamefont {Watanabe}, \citenamefont {Grigorieva}, \citenamefont {Polini}, \citenamefont {Goltsman}, \citenamefont {Geim},\ and\ \citenamefont {Fedorov}}]{Bandurin_resonant}%
  \BibitemOpen
  \bibfield  {author} {\bibinfo {author} {\bibfnamefont {Denis~A.}\ \bibnamefont {Bandurin}}, \bibinfo {author} {\bibfnamefont {Dmitry}\ \bibnamefont {Svintsov}}, \bibinfo {author} {\bibfnamefont {Igor}\ \bibnamefont {Gayduchenko}}, \bibinfo {author} {\bibfnamefont {Shuigang~G.}\ \bibnamefont {Xu}}, \bibinfo {author} {\bibfnamefont {Alessandro}\ \bibnamefont {Principi}}, \bibinfo {author} {\bibfnamefont {Maxim}\ \bibnamefont {Moskotin}}, \bibinfo {author} {\bibfnamefont {Ivan}\ \bibnamefont {Tretyakov}}, \bibinfo {author} {\bibfnamefont {Denis}\ \bibnamefont {Yagodkin}}, \bibinfo {author} {\bibfnamefont {Sergey}\ \bibnamefont {Zhukov}}, \bibinfo {author} {\bibfnamefont {Takashi}\ \bibnamefont {Taniguchi}}, \bibinfo {author} {\bibfnamefont {Kenji}\ \bibnamefont {Watanabe}}, \bibinfo {author} {\bibfnamefont {Irina~V.}\ \bibnamefont {Grigorieva}}, \bibinfo {author} {\bibfnamefont {Marco}\ \bibnamefont {Polini}}, \bibinfo {author} {\bibfnamefont {Gregory~N.}\ \bibnamefont {Goltsman}}, \bibinfo {author}
  {\bibfnamefont {Andre~K.}\ \bibnamefont {Geim}}, \ and\ \bibinfo {author} {\bibfnamefont {Georgy}\ \bibnamefont {Fedorov}},\ }\bibfield  {title} {\enquote {\bibinfo {title} {{Resonant terahertz detection using graphene plasmons}},}\ }\href {\doibase 10.1038/s41467-018-07848-w} {\bibfield  {journal} {\bibinfo  {journal} {Nature Communications}\ }\textbf {\bibinfo {volume} {9}},\ \bibinfo {pages} {5392} (\bibinfo {year} {2018})}\BibitemShut {NoStop}%
\bibitem [{\citenamefont {Muravev}\ \emph {et~al.}(2016)\citenamefont {Muravev}, \citenamefont {Fortunatov}, \citenamefont {Dremin},\ and\ \citenamefont {Kukushkin}}]{Muravev_interferometer}%
  \BibitemOpen
  \bibfield  {author} {\bibinfo {author} {\bibfnamefont {V.~M.}\ \bibnamefont {Muravev}}, \bibinfo {author} {\bibfnamefont {A.~A.}\ \bibnamefont {Fortunatov}}, \bibinfo {author} {\bibfnamefont {A.~A.}\ \bibnamefont {Dremin}}, \ and\ \bibinfo {author} {\bibfnamefont {I.~V.}\ \bibnamefont {Kukushkin}},\ }\bibfield  {title} {\enquote {\bibinfo {title} {{Plasmonic interferometer for spectroscopy of microwave radiation}},}\ }\href {\doibase 10.1134/S0021364016060084} {\bibfield  {journal} {\bibinfo  {journal} {JETP Letters}\ }\textbf {\bibinfo {volume} {103}},\ \bibinfo {pages} {380--384} (\bibinfo {year} {2016})}\BibitemShut {NoStop}%
\bibitem [{\citenamefont {Rodrigo}\ \emph {et~al.}(2015)\citenamefont {Rodrigo}, \citenamefont {Limaj}, \citenamefont {Janner}, \citenamefont {Etezadi}, \citenamefont {{Garc{\'{i}}a De Abajo}}, \citenamefont {Pruneri},\ and\ \citenamefont {Altug}}]{Rodrigo_biosensing}%
  \BibitemOpen
  \bibfield  {author} {\bibinfo {author} {\bibfnamefont {Daniel}\ \bibnamefont {Rodrigo}}, \bibinfo {author} {\bibfnamefont {Odeta}\ \bibnamefont {Limaj}}, \bibinfo {author} {\bibfnamefont {Davide}\ \bibnamefont {Janner}}, \bibinfo {author} {\bibfnamefont {Dordaneh}\ \bibnamefont {Etezadi}}, \bibinfo {author} {\bibfnamefont {F.~Javier}\ \bibnamefont {{Garc{\'{i}}a De Abajo}}}, \bibinfo {author} {\bibfnamefont {Valerio}\ \bibnamefont {Pruneri}}, \ and\ \bibinfo {author} {\bibfnamefont {Hatice}\ \bibnamefont {Altug}},\ }\bibfield  {title} {\enquote {\bibinfo {title} {{Mid-infrared plasmonic biosensing with graphene}},}\ }\href {\doibase 10.1126/science.aab2051} {\bibfield  {journal} {\bibinfo  {journal} {Science}\ }\textbf {\bibinfo {volume} {349}},\ \bibinfo {pages} {165--168} (\bibinfo {year} {2015})}\BibitemShut {NoStop}%
\bibitem [{\citenamefont {Bylinkin}\ \emph {et~al.}(2024)\citenamefont {Bylinkin}, \citenamefont {Castilla}, \citenamefont {Slipchenko}, \citenamefont {Domina}, \citenamefont {Calavalle}, \citenamefont {Pusapati}, \citenamefont {Autore}, \citenamefont {Casanova}, \citenamefont {Hueso}, \citenamefont {Mart{\'{i}}n-Moreno}, \citenamefont {Nikitin}, \citenamefont {Koppens},\ and\ \citenamefont {Hillenbrand}}]{Bylinkin2024}%
  \BibitemOpen
  \bibfield  {author} {\bibinfo {author} {\bibfnamefont {Andrei}\ \bibnamefont {Bylinkin}}, \bibinfo {author} {\bibfnamefont {Sebasti{\'{a}}n}\ \bibnamefont {Castilla}}, \bibinfo {author} {\bibfnamefont {Tetiana~M.}\ \bibnamefont {Slipchenko}}, \bibinfo {author} {\bibfnamefont {Kateryna}\ \bibnamefont {Domina}}, \bibinfo {author} {\bibfnamefont {Francesco}\ \bibnamefont {Calavalle}}, \bibinfo {author} {\bibfnamefont {Varun-Varma}\ \bibnamefont {Pusapati}}, \bibinfo {author} {\bibfnamefont {Marta}\ \bibnamefont {Autore}}, \bibinfo {author} {\bibfnamefont {F{\`{e}}lix}\ \bibnamefont {Casanova}}, \bibinfo {author} {\bibfnamefont {Luis~E.}\ \bibnamefont {Hueso}}, \bibinfo {author} {\bibfnamefont {Luis}\ \bibnamefont {Mart{\'{i}}n-Moreno}}, \bibinfo {author} {\bibfnamefont {Alexey~Y.}\ \bibnamefont {Nikitin}}, \bibinfo {author} {\bibfnamefont {Frank H.~L.}\ \bibnamefont {Koppens}}, \ and\ \bibinfo {author} {\bibfnamefont {Rainer}\ \bibnamefont {Hillenbrand}},\ }\bibfield  {title} {\enquote {\bibinfo {title}
  {{On-chip phonon-enhanced IR near-field detection of molecular vibrations}},}\ }\href {\doibase 10.1038/s41467-024-53182-9} {\bibfield  {journal} {\bibinfo  {journal} {Nature Communications}\ }\textbf {\bibinfo {volume} {15}},\ \bibinfo {pages} {8907} (\bibinfo {year} {2024})}\BibitemShut {NoStop}%
\bibitem [{\citenamefont {Li}\ \emph {et~al.}(2019)\citenamefont {Li}, \citenamefont {Ferreyra}, \citenamefont {Swan},\ and\ \citenamefont {Paiella}}]{Graphene_thermal_emission}%
  \BibitemOpen
  \bibfield  {author} {\bibinfo {author} {\bibfnamefont {Yuyu}\ \bibnamefont {Li}}, \bibinfo {author} {\bibfnamefont {Pablo}\ \bibnamefont {Ferreyra}}, \bibinfo {author} {\bibfnamefont {Anna~K.}\ \bibnamefont {Swan}}, \ and\ \bibinfo {author} {\bibfnamefont {Roberto}\ \bibnamefont {Paiella}},\ }\bibfield  {title} {\enquote {\bibinfo {title} {{Current-Driven Terahertz Light Emission from Graphene Plasmonic Oscillations}},}\ }\href {\doibase 10.1021/acsphotonics.9b01037} {\bibfield  {journal} {\bibinfo  {journal} {ACS Photonics}\ }\textbf {\bibinfo {volume} {6}},\ \bibinfo {pages} {2562--2569} (\bibinfo {year} {2019})}\BibitemShut {NoStop}%
\bibitem [{\citenamefont {Dyakonov}\ and\ \citenamefont {Shur}(1993)}]{Dyakonov1993}%
  \BibitemOpen
  \bibfield  {author} {\bibinfo {author} {\bibfnamefont {Michael}\ \bibnamefont {Dyakonov}}\ and\ \bibinfo {author} {\bibfnamefont {Michael}\ \bibnamefont {Shur}},\ }\bibfield  {title} {\enquote {\bibinfo {title} {{Shallow water analogy for a ballistic field effect transistor: New mechanism of plasma wave generation by dc current}},}\ }\href {\doibase 10.1103/PhysRevLett.71.2465} {\bibfield  {journal} {\bibinfo  {journal} {Physical Review Letters}\ }\textbf {\bibinfo {volume} {71}},\ \bibinfo {pages} {2465--2468} (\bibinfo {year} {1993})}\BibitemShut {NoStop}%
\bibitem [{\citenamefont {{El Fatimy}}\ \emph {et~al.}(2010)\citenamefont {{El Fatimy}}, \citenamefont {Dyakonova}, \citenamefont {Meziani}, \citenamefont {Otsuji}, \citenamefont {Knap}, \citenamefont {Vandenbrouk}, \citenamefont {Madjour}, \citenamefont {Th{\'{e}}ron}, \citenamefont {Gaquiere}, \citenamefont {Poisson}, \citenamefont {Delage}, \citenamefont {Prystawko},\ and\ \citenamefont {Skierbiszewski}}]{ElFatimy2010}%
  \BibitemOpen
  \bibfield  {author} {\bibinfo {author} {\bibfnamefont {A.}~\bibnamefont {{El Fatimy}}}, \bibinfo {author} {\bibfnamefont {N.}~\bibnamefont {Dyakonova}}, \bibinfo {author} {\bibfnamefont {Y.}~\bibnamefont {Meziani}}, \bibinfo {author} {\bibfnamefont {T.}~\bibnamefont {Otsuji}}, \bibinfo {author} {\bibfnamefont {W.}~\bibnamefont {Knap}}, \bibinfo {author} {\bibfnamefont {S.}~\bibnamefont {Vandenbrouk}}, \bibinfo {author} {\bibfnamefont {K.}~\bibnamefont {Madjour}}, \bibinfo {author} {\bibfnamefont {D.}~\bibnamefont {Th{\'{e}}ron}}, \bibinfo {author} {\bibfnamefont {C.}~\bibnamefont {Gaquiere}}, \bibinfo {author} {\bibfnamefont {M.~A.}\ \bibnamefont {Poisson}}, \bibinfo {author} {\bibfnamefont {S.}~\bibnamefont {Delage}}, \bibinfo {author} {\bibfnamefont {P.}~\bibnamefont {Prystawko}}, \ and\ \bibinfo {author} {\bibfnamefont {C.}~\bibnamefont {Skierbiszewski}},\ }\bibfield  {title} {\enquote {\bibinfo {title} {{AlGaN/GaN high electron mobility transistors as a voltage-tunable room temperature terahertz
  sources}},}\ }\href {\doibase 10.1063/1.3291101} {\bibfield  {journal} {\bibinfo  {journal} {Journal of Applied Physics}\ }\textbf {\bibinfo {volume} {107}},\ \bibinfo {pages} {024504} (\bibinfo {year} {2010})}\BibitemShut {NoStop}%
\bibitem [{\citenamefont {Hwang}\ and\ \citenamefont {{Das Sarma}}(2007)}]{Hwang2007a}%
  \BibitemOpen
  \bibfield  {author} {\bibinfo {author} {\bibfnamefont {E.~H.}\ \bibnamefont {Hwang}}\ and\ \bibinfo {author} {\bibfnamefont {S.}~\bibnamefont {{Das Sarma}}},\ }\bibfield  {title} {\enquote {\bibinfo {title} {{Dielectric function, screening, and plasmons in two-dimensional graphene}},}\ }\href {\doibase 10.1103/PhysRevB.75.205418} {\bibfield  {journal} {\bibinfo  {journal} {Physical Review B}\ }\textbf {\bibinfo {volume} {75}},\ \bibinfo {pages} {205418} (\bibinfo {year} {2007})}\BibitemShut {NoStop}%
\bibitem [{\citenamefont {Dias}\ and\ \citenamefont {{Garc{\'{i}}a De Abajo}}(2019)}]{Limits_to_plasmon_coupling}%
  \BibitemOpen
  \bibfield  {author} {\bibinfo {author} {\bibfnamefont {Eduardo~J.C.}\ \bibnamefont {Dias}}\ and\ \bibinfo {author} {\bibfnamefont {F.~Javier}\ \bibnamefont {{Garc{\'{i}}a De Abajo}}},\ }\bibfield  {title} {\enquote {\bibinfo {title} {{Fundamental Limits to the Coupling between Light and 2D Polaritons by Small Scatterers}},}\ }\href {\doibase 10.1021/acsnano.8b09283} {\bibfield  {journal} {\bibinfo  {journal} {ACS Nano}\ }\textbf {\bibinfo {volume} {13}},\ \bibinfo {pages} {5184--5197} (\bibinfo {year} {2019})}\BibitemShut {NoStop}%
\bibitem [{\citenamefont {Chaplik}(1985)}]{Chaplik_emission_absorption}%
  \BibitemOpen
  \bibfield  {author} {\bibinfo {author} {\bibfnamefont {A.V.}\ \bibnamefont {Chaplik}},\ }\bibfield  {title} {\enquote {\bibinfo {title} {{Absorption and emission of electromagnetic waves by two-dimensional plasmons}},}\ }\href {\doibase 10.1016/0167-5729(85)90010-X} {\bibfield  {journal} {\bibinfo  {journal} {Surface Science Reports}\ }\textbf {\bibinfo {volume} {5}},\ \bibinfo {pages} {289--335} (\bibinfo {year} {1985})}\BibitemShut {NoStop}%
\bibitem [{\citenamefont {Fateev}\ \emph {et~al.}(2010)\citenamefont {Fateev}, \citenamefont {Popov},\ and\ \citenamefont {Shur}}]{Fateev_transormation}%
  \BibitemOpen
  \bibfield  {author} {\bibinfo {author} {\bibfnamefont {D.~V.}\ \bibnamefont {Fateev}}, \bibinfo {author} {\bibfnamefont {V.~V.}\ \bibnamefont {Popov}}, \ and\ \bibinfo {author} {\bibfnamefont {M.~S.}\ \bibnamefont {Shur}},\ }\bibfield  {title} {\enquote {\bibinfo {title} {{Transformation of the plasmon spectrum in a grating-gate transistor structure with spatially modulated two-dimensional electron channel}},}\ }\href {\doibase 10.1134/S1063782610110059} {\bibfield  {journal} {\bibinfo  {journal} {Semiconductors}\ }\textbf {\bibinfo {volume} {44}},\ \bibinfo {pages} {1406--1413} (\bibinfo {year} {2010})}\BibitemShut {NoStop}%
\bibitem [{\citenamefont {McLeod}\ \emph {et~al.}(2014)\citenamefont {McLeod}, \citenamefont {Kelly}, \citenamefont {Goldflam}, \citenamefont {Gainsforth}, \citenamefont {Westphal}, \citenamefont {Dominguez}, \citenamefont {Thiemens}, \citenamefont {Fogler},\ and\ \citenamefont {Basov}}]{McLeod2014}%
  \BibitemOpen
  \bibfield  {author} {\bibinfo {author} {\bibfnamefont {Alexander~S.}\ \bibnamefont {McLeod}}, \bibinfo {author} {\bibfnamefont {P.}~\bibnamefont {Kelly}}, \bibinfo {author} {\bibfnamefont {M.~D.}\ \bibnamefont {Goldflam}}, \bibinfo {author} {\bibfnamefont {Z.}~\bibnamefont {Gainsforth}}, \bibinfo {author} {\bibfnamefont {A.~J.}\ \bibnamefont {Westphal}}, \bibinfo {author} {\bibfnamefont {Gerardo}\ \bibnamefont {Dominguez}}, \bibinfo {author} {\bibfnamefont {Mark~H.}\ \bibnamefont {Thiemens}}, \bibinfo {author} {\bibfnamefont {Michael~M.}\ \bibnamefont {Fogler}}, \ and\ \bibinfo {author} {\bibfnamefont {D.~N.}\ \bibnamefont {Basov}},\ }\bibfield  {title} {\enquote {\bibinfo {title} {{Model for quantitative tip-enhanced spectroscopy and the extraction of nanoscale-resolved optical constants}},}\ }\href {\doibase 10.1103/PhysRevB.90.085136} {\bibfield  {journal} {\bibinfo  {journal} {Physical Review B}\ }\textbf {\bibinfo {volume} {90}},\ \bibinfo {pages} {085136} (\bibinfo {year} {2014})}\BibitemShut
  {NoStop}%
\bibitem [{\citenamefont {Woessner}\ \emph {et~al.}(2017{\natexlab{b}})\citenamefont {Woessner}, \citenamefont {Gao}, \citenamefont {Torre}, \citenamefont {Lundeberg}, \citenamefont {Tan}, \citenamefont {Watanabe}, \citenamefont {Taniguchi}, \citenamefont {Hillenbrand}, \citenamefont {Hone}, \citenamefont {Polini},\ and\ \citenamefont {Koppens}}]{Woessner_phase_shifter}%
  \BibitemOpen
  \bibfield  {author} {\bibinfo {author} {\bibfnamefont {Achim}\ \bibnamefont {Woessner}}, \bibinfo {author} {\bibfnamefont {Yuanda}\ \bibnamefont {Gao}}, \bibinfo {author} {\bibfnamefont {Iacopo}\ \bibnamefont {Torre}}, \bibinfo {author} {\bibfnamefont {Mark~B.}\ \bibnamefont {Lundeberg}}, \bibinfo {author} {\bibfnamefont {Cheng}\ \bibnamefont {Tan}}, \bibinfo {author} {\bibfnamefont {Kenji}\ \bibnamefont {Watanabe}}, \bibinfo {author} {\bibfnamefont {Takashi}\ \bibnamefont {Taniguchi}}, \bibinfo {author} {\bibfnamefont {Rainer}\ \bibnamefont {Hillenbrand}}, \bibinfo {author} {\bibfnamefont {James}\ \bibnamefont {Hone}}, \bibinfo {author} {\bibfnamefont {Marco}\ \bibnamefont {Polini}}, \ and\ \bibinfo {author} {\bibfnamefont {Frank~H.L.}\ \bibnamefont {Koppens}},\ }\bibfield  {title} {\enquote {\bibinfo {title} {{Electrical 2$\pi$ phase control of infrared light in a 350-nm footprint using graphene plasmons}},}\ }\href {\doibase 10.1038/nphoton.2017.98} {\bibfield  {journal} {\bibinfo  {journal} {Nature
  Photonics}\ }\textbf {\bibinfo {volume} {11}},\ \bibinfo {pages} {421--424} (\bibinfo {year} {2017}{\natexlab{b}})}\BibitemShut {NoStop}%
\bibitem [{\citenamefont {Ni}\ \emph {et~al.}(2018)\citenamefont {Ni}, \citenamefont {McLeod}, \citenamefont {Sun}, \citenamefont {Wang}, \citenamefont {Xiong}, \citenamefont {Post}, \citenamefont {Sunku}, \citenamefont {Jiang}, \citenamefont {Hone}, \citenamefont {Dean}, \citenamefont {Fogler},\ and\ \citenamefont {Basov}}]{Ni_limits_plasmonics}%
  \BibitemOpen
  \bibfield  {author} {\bibinfo {author} {\bibfnamefont {G.~X.}\ \bibnamefont {Ni}}, \bibinfo {author} {\bibfnamefont {A.~S.}\ \bibnamefont {McLeod}}, \bibinfo {author} {\bibfnamefont {Z.}~\bibnamefont {Sun}}, \bibinfo {author} {\bibfnamefont {L.}~\bibnamefont {Wang}}, \bibinfo {author} {\bibfnamefont {L.}~\bibnamefont {Xiong}}, \bibinfo {author} {\bibfnamefont {K.~W.}\ \bibnamefont {Post}}, \bibinfo {author} {\bibfnamefont {S.~S.}\ \bibnamefont {Sunku}}, \bibinfo {author} {\bibfnamefont {B.-Y.}\ \bibnamefont {Jiang}}, \bibinfo {author} {\bibfnamefont {J.}~\bibnamefont {Hone}}, \bibinfo {author} {\bibfnamefont {C.~R.}\ \bibnamefont {Dean}}, \bibinfo {author} {\bibfnamefont {M.~M.}\ \bibnamefont {Fogler}}, \ and\ \bibinfo {author} {\bibfnamefont {D.~N.}\ \bibnamefont {Basov}},\ }\bibfield  {title} {\enquote {\bibinfo {title} {{Fundamental limits to graphene plasmonics}},}\ }\href {\doibase 10.1038/s41586-018-0136-9} {\bibfield  {journal} {\bibinfo  {journal} {Nature}\ }\textbf {\bibinfo {volume} {557}},\
  \bibinfo {pages} {530--533} (\bibinfo {year} {2018})}\BibitemShut {NoStop}%
\bibitem [{\citenamefont {Castilla}\ \emph {et~al.}(2019)\citenamefont {Castilla}, \citenamefont {Terr{\'{e}}s}, \citenamefont {Autore}, \citenamefont {Viti}, \citenamefont {Li}, \citenamefont {Nikitin}, \citenamefont {Vangelidis}, \citenamefont {Watanabe}, \citenamefont {Taniguchi}, \citenamefont {Lidorikis}, \citenamefont {Vitiello}, \citenamefont {Hillenbrand}, \citenamefont {Tielrooij},\ and\ \citenamefont {Koppens}}]{Castilla_fast_THz}%
  \BibitemOpen
  \bibfield  {author} {\bibinfo {author} {\bibfnamefont {Sebasti{\'{a}}n}\ \bibnamefont {Castilla}}, \bibinfo {author} {\bibfnamefont {Bernat}\ \bibnamefont {Terr{\'{e}}s}}, \bibinfo {author} {\bibfnamefont {Marta}\ \bibnamefont {Autore}}, \bibinfo {author} {\bibfnamefont {Leonardo}\ \bibnamefont {Viti}}, \bibinfo {author} {\bibfnamefont {Jian}\ \bibnamefont {Li}}, \bibinfo {author} {\bibfnamefont {Alexey~Y.}\ \bibnamefont {Nikitin}}, \bibinfo {author} {\bibfnamefont {Ioannis}\ \bibnamefont {Vangelidis}}, \bibinfo {author} {\bibfnamefont {Kenji}\ \bibnamefont {Watanabe}}, \bibinfo {author} {\bibfnamefont {Takashi}\ \bibnamefont {Taniguchi}}, \bibinfo {author} {\bibfnamefont {Elefterios}\ \bibnamefont {Lidorikis}}, \bibinfo {author} {\bibfnamefont {Miriam~S.}\ \bibnamefont {Vitiello}}, \bibinfo {author} {\bibfnamefont {Rainer}\ \bibnamefont {Hillenbrand}}, \bibinfo {author} {\bibfnamefont {Klaas-Jan}\ \bibnamefont {Tielrooij}}, \ and\ \bibinfo {author} {\bibfnamefont {Frank~H.L.}\ \bibnamefont {Koppens}},\
  }\bibfield  {title} {\enquote {\bibinfo {title} {{Fast and Sensitive Terahertz Detection Using an Antenna-Integrated Graphene pn Junction}},}\ }\href {\doibase 10.1021/acs.nanolett.8b04171} {\bibfield  {journal} {\bibinfo  {journal} {Nano Letters}\ }\textbf {\bibinfo {volume} {19}},\ \bibinfo {pages} {2765--2773} (\bibinfo {year} {2019})}\BibitemShut {NoStop}%
\bibitem [{\citenamefont {Olbrich}\ \emph {et~al.}(2016)\citenamefont {Olbrich}, \citenamefont {Kamann}, \citenamefont {K{\"{o}}nig}, \citenamefont {Munzert}, \citenamefont {Tutsch}, \citenamefont {Eroms}, \citenamefont {Weiss}, \citenamefont {Liu}, \citenamefont {Golub}, \citenamefont {Ivchenko}, \citenamefont {Popov}, \citenamefont {Fateev}, \citenamefont {Mashinsky}, \citenamefont {Fromm}, \citenamefont {Seyller},\ and\ \citenamefont {Ganichev}}]{Olbrich2016}%
  \BibitemOpen
  \bibfield  {author} {\bibinfo {author} {\bibfnamefont {P.}~\bibnamefont {Olbrich}}, \bibinfo {author} {\bibfnamefont {J.}~\bibnamefont {Kamann}}, \bibinfo {author} {\bibfnamefont {M.}~\bibnamefont {K{\"{o}}nig}}, \bibinfo {author} {\bibfnamefont {J.}~\bibnamefont {Munzert}}, \bibinfo {author} {\bibfnamefont {L.}~\bibnamefont {Tutsch}}, \bibinfo {author} {\bibfnamefont {J.}~\bibnamefont {Eroms}}, \bibinfo {author} {\bibfnamefont {D.}~\bibnamefont {Weiss}}, \bibinfo {author} {\bibfnamefont {Ming-Hao}\ \bibnamefont {Liu}}, \bibinfo {author} {\bibfnamefont {L.~E.}\ \bibnamefont {Golub}}, \bibinfo {author} {\bibfnamefont {E.~L.}\ \bibnamefont {Ivchenko}}, \bibinfo {author} {\bibfnamefont {V.~V.}\ \bibnamefont {Popov}}, \bibinfo {author} {\bibfnamefont {D.~V.}\ \bibnamefont {Fateev}}, \bibinfo {author} {\bibfnamefont {K.~V.}\ \bibnamefont {Mashinsky}}, \bibinfo {author} {\bibfnamefont {F.}~\bibnamefont {Fromm}}, \bibinfo {author} {\bibfnamefont {Th}~\bibnamefont {Seyller}}, \ and\ \bibinfo {author} {\bibfnamefont
  {S.~D.}\ \bibnamefont {Ganichev}},\ }\bibfield  {title} {\enquote {\bibinfo {title} {{Terahertz ratchet effects in graphene with a lateral superlattice}},}\ }\href {\doibase 10.1103/PhysRevB.93.075422} {\bibfield  {journal} {\bibinfo  {journal} {Physical Review B}\ }\textbf {\bibinfo {volume} {93}},\ \bibinfo {pages} {075422} (\bibinfo {year} {2016})}\BibitemShut {NoStop}%
\bibitem [{\citenamefont {Sydoruk}\ \emph {et~al.}(2015)\citenamefont {Sydoruk}, \citenamefont {Choonee},\ and\ \citenamefont {Dyer}}]{Sydoruk_gate_edge}%
  \BibitemOpen
  \bibfield  {author} {\bibinfo {author} {\bibfnamefont {Oleksiy}\ \bibnamefont {Sydoruk}}, \bibinfo {author} {\bibfnamefont {Kaushal}\ \bibnamefont {Choonee}}, \ and\ \bibinfo {author} {\bibfnamefont {Gregory~C.}\ \bibnamefont {Dyer}},\ }\bibfield  {title} {\enquote {\bibinfo {title} {{Transmission and Reflection of Terahertz Plasmons in Two-Dimensional Plasmonic Devices}},}\ }\href {\doibase 10.1109/TTHZ.2015.2405919} {\bibfield  {journal} {\bibinfo  {journal} {IEEE Transactions on Terahertz Science and Technology}\ }\textbf {\bibinfo {volume} {5}},\ \bibinfo {pages} {486--496} (\bibinfo {year} {2015})}\BibitemShut {NoStop}%
\bibitem [{\citenamefont {Bylinkin}\ \emph {et~al.}(2019)\citenamefont {Bylinkin}, \citenamefont {Titova}, \citenamefont {Mikheev}, \citenamefont {Zhukova}, \citenamefont {Zhukov}, \citenamefont {Belyanchikov}, \citenamefont {Kashchenko}, \citenamefont {Miakonkikh},\ and\ \citenamefont {Svintsov}}]{Bylinkin_tight_binding}%
  \BibitemOpen
  \bibfield  {author} {\bibinfo {author} {\bibfnamefont {Andrey}\ \bibnamefont {Bylinkin}}, \bibinfo {author} {\bibfnamefont {Elena}\ \bibnamefont {Titova}}, \bibinfo {author} {\bibfnamefont {Vitaly}\ \bibnamefont {Mikheev}}, \bibinfo {author} {\bibfnamefont {Elena}\ \bibnamefont {Zhukova}}, \bibinfo {author} {\bibfnamefont {Sergey}\ \bibnamefont {Zhukov}}, \bibinfo {author} {\bibfnamefont {Mikhail}\ \bibnamefont {Belyanchikov}}, \bibinfo {author} {\bibfnamefont {Mikhail}\ \bibnamefont {Kashchenko}}, \bibinfo {author} {\bibfnamefont {Andrew}\ \bibnamefont {Miakonkikh}}, \ and\ \bibinfo {author} {\bibfnamefont {Dmitry}\ \bibnamefont {Svintsov}},\ }\bibfield  {title} {\enquote {\bibinfo {title} {{Tight-Binding Terahertz Plasmons in Chemical-Vapor-Deposited Graphene}},}\ }\href {\doibase 10.1103/PhysRevApplied.11.054017} {\bibfield  {journal} {\bibinfo  {journal} {Physical Review Applied}\ }\textbf {\bibinfo {volume} {11}},\ \bibinfo {pages} {054017} (\bibinfo {year} {2019})}\BibitemShut {NoStop}%
\bibitem [{\citenamefont {Sai}\ \emph {et~al.}(2023)\citenamefont {Sai}, \citenamefont {Korotyeyev}, \citenamefont {Dub}, \citenamefont {S\l{}owikowski}, \citenamefont {Filipiak}, \citenamefont {But}, \citenamefont {Ivonyak}, \citenamefont {Sakowicz}, \citenamefont {Lyaschuk}, \citenamefont {Kukhtaruk}, \citenamefont {Cywi\ifmmode~\acute{n}\else \'{n}\fi{}ski},\ and\ \citenamefont {Knap}}]{Knap_crystals}%
  \BibitemOpen
  \bibfield  {author} {\bibinfo {author} {\bibfnamefont {P.}~\bibnamefont {Sai}}, \bibinfo {author} {\bibfnamefont {V.~V.}\ \bibnamefont {Korotyeyev}}, \bibinfo {author} {\bibfnamefont {M.}~\bibnamefont {Dub}}, \bibinfo {author} {\bibfnamefont {M.}~\bibnamefont {S\l{}owikowski}}, \bibinfo {author} {\bibfnamefont {M.}~\bibnamefont {Filipiak}}, \bibinfo {author} {\bibfnamefont {D.~B.}\ \bibnamefont {But}}, \bibinfo {author} {\bibfnamefont {Yu.}\ \bibnamefont {Ivonyak}}, \bibinfo {author} {\bibfnamefont {M.}~\bibnamefont {Sakowicz}}, \bibinfo {author} {\bibfnamefont {Yu.~M.}\ \bibnamefont {Lyaschuk}}, \bibinfo {author} {\bibfnamefont {S.~M.}\ \bibnamefont {Kukhtaruk}}, \bibinfo {author} {\bibfnamefont {G.}~\bibnamefont {Cywi\ifmmode~\acute{n}\else \'{n}\fi{}ski}}, \ and\ \bibinfo {author} {\bibfnamefont {W.}~\bibnamefont {Knap}},\ }\bibfield  {title} {\enquote {\bibinfo {title} {Electrical tuning of terahertz plasmonic crystal phases},}\ }\href {\doibase 10.1103/PhysRevX.13.041003} {\bibfield  {journal}
  {\bibinfo  {journal} {Phys. Rev. X}\ }\textbf {\bibinfo {volume} {13}},\ \bibinfo {pages} {041003} (\bibinfo {year} {2023})}\BibitemShut {NoStop}%
\bibitem [{\citenamefont {Shuvaev}\ \emph {et~al.}(2022)\citenamefont {Shuvaev}, \citenamefont {Dzhikirba}, \citenamefont {Astrakhantseva}, \citenamefont {Gusikhin}, \citenamefont {Kukushkin},\ and\ \citenamefont {Muravev}}]{Muravev_grating}%
  \BibitemOpen
  \bibfield  {author} {\bibinfo {author} {\bibfnamefont {A.}~\bibnamefont {Shuvaev}}, \bibinfo {author} {\bibfnamefont {K.~R.}\ \bibnamefont {Dzhikirba}}, \bibinfo {author} {\bibfnamefont {A.~S.}\ \bibnamefont {Astrakhantseva}}, \bibinfo {author} {\bibfnamefont {P.~A.}\ \bibnamefont {Gusikhin}}, \bibinfo {author} {\bibfnamefont {I.~V.}\ \bibnamefont {Kukushkin}}, \ and\ \bibinfo {author} {\bibfnamefont {V.~M.}\ \bibnamefont {Muravev}},\ }\bibfield  {title} {\enquote {\bibinfo {title} {Plasmonic metasurface created by a grating of two-dimensional electron strips on a substrate},}\ }\href {\doibase 10.1103/PhysRevB.106.L161411} {\bibfield  {journal} {\bibinfo  {journal} {Phys. Rev. B}\ }\textbf {\bibinfo {volume} {106}},\ \bibinfo {pages} {L161411} (\bibinfo {year} {2022})}\BibitemShut {NoStop}%
\bibitem [{\citenamefont {Senior}(1952)}]{Senior1952}%
  \BibitemOpen
  \bibfield  {author} {\bibinfo {author} {\bibfnamefont {T.~B.A.}\ \bibnamefont {Senior}},\ }\bibfield  {title} {\enquote {\bibinfo {title} {{Diffraction by a semi-infinite metallic sheet}},}\ }\href {\doibase 10.1098/rspa.1952.0137} {\bibfield  {journal} {\bibinfo  {journal} {Proceedings of the Royal Society of London. Series A. Mathematical and Physical Sciences}\ }\textbf {\bibinfo {volume} {213}},\ \bibinfo {pages} {436--458} (\bibinfo {year} {1952})}\BibitemShut {NoStop}%
\bibitem [{\citenamefont {Nussenzveig}\ and\ \citenamefont {Lighthill}(1959)}]{Nussenzveig_WG_termination}%
  \BibitemOpen
  \bibfield  {author} {\bibinfo {author} {\bibfnamefont {H.~M.}\ \bibnamefont {Nussenzveig}}\ and\ \bibinfo {author} {\bibfnamefont {Michael~James}\ \bibnamefont {Lighthill}},\ }\bibfield  {title} {\enquote {\bibinfo {title} {Solution of a diffraction problem - solution of a diffraction problem. i the wide double wedge},}\ }\href {\doibase 10.1098/rsta.1959.0012} {\bibfield  {journal} {\bibinfo  {journal} {Philosophical Transactions of the Royal Society of London. Series A, Mathematical and Physical Sciences}\ }\textbf {\bibinfo {volume} {252}},\ \bibinfo {pages} {1--30} (\bibinfo {year} {1959})}\BibitemShut {NoStop}%
\bibitem [{\citenamefont {Kay}(1959)}]{Kay_reactance_discontinuity}%
  \BibitemOpen
  \bibfield  {author} {\bibinfo {author} {\bibfnamefont {A.}~\bibnamefont {Kay}},\ }\bibfield  {title} {\enquote {\bibinfo {title} {{Scattering of a surface wave by a discontinuity in reactance}},}\ }\href {\doibase 10.1109/TAP.1959.1144635} {\bibfield  {journal} {\bibinfo  {journal} {IRE Transactions on Antennas and Propagation}\ }\textbf {\bibinfo {volume} {7}},\ \bibinfo {pages} {22--31} (\bibinfo {year} {1959})}\BibitemShut {NoStop}%
\bibitem [{\citenamefont {Volkov}\ and\ \citenamefont {Mikhailov}(1988)}]{volkov1988edge}%
  \BibitemOpen
  \bibfield  {author} {\bibinfo {author} {\bibfnamefont {VA}~\bibnamefont {Volkov}}\ and\ \bibinfo {author} {\bibfnamefont {Sergey~A}\ \bibnamefont {Mikhailov}},\ }\bibfield  {title} {\enquote {\bibinfo {title} {Edge magnetoplasmons: low frequency weakly damped excitations in inhomogeneous two-dimensional electron systems},}\ }\href@noop {} {\bibfield  {journal} {\bibinfo  {journal} {Sov. Phys. JETP}\ }\textbf {\bibinfo {volume} {67}},\ \bibinfo {pages} {1639--1653} (\bibinfo {year} {1988})}\BibitemShut {NoStop}%
\bibitem [{\citenamefont {Margetis}\ \emph {et~al.}(2017)\citenamefont {Margetis}, \citenamefont {Maier},\ and\ \citenamefont {Luskin}}]{Margetis2017}%
  \BibitemOpen
  \bibfield  {author} {\bibinfo {author} {\bibfnamefont {Dionisios}\ \bibnamefont {Margetis}}, \bibinfo {author} {\bibfnamefont {Matthias}\ \bibnamefont {Maier}}, \ and\ \bibinfo {author} {\bibfnamefont {Mitchell}\ \bibnamefont {Luskin}},\ }\bibfield  {title} {\enquote {\bibinfo {title} {{On the Wiener–Hopf Method for Surface Plasmons: Diffraction from Semiinfinite Metamaterial Sheet}},}\ }\href {\doibase 10.1111/sapm.12180} {\bibfield  {journal} {\bibinfo  {journal} {Studies in Applied Mathematics}\ }\textbf {\bibinfo {volume} {139}},\ \bibinfo {pages} {599--625} (\bibinfo {year} {2017})}\BibitemShut {NoStop}%
\bibitem [{\citenamefont {Zhang}\ \emph {et~al.}(2014)\citenamefont {Zhang}, \citenamefont {Fu},\ and\ \citenamefont {Yang}}]{Zhang_Wiener_Hopf}%
  \BibitemOpen
  \bibfield  {author} {\bibinfo {author} {\bibfnamefont {Lei}\ \bibnamefont {Zhang}}, \bibinfo {author} {\bibfnamefont {Xiu~Li}\ \bibnamefont {Fu}}, \ and\ \bibinfo {author} {\bibfnamefont {Jun~Zhong}\ \bibnamefont {Yang}},\ }\bibfield  {title} {\enquote {\bibinfo {title} {{Excitation of propagating plasmons in semi-infinite graphene layer by free space photons}},}\ }\href {\doibase 10.1088/0253-6102/61/6/14} {\bibfield  {journal} {\bibinfo  {journal} {Communications in Theoretical Physics}\ }\textbf {\bibinfo {volume} {61}},\ \bibinfo {pages} {751--754} (\bibinfo {year} {2014})}\BibitemShut {NoStop}%
\bibitem [{\citenamefont {Nikulin}\ \emph {et~al.}(2021)\citenamefont {Nikulin}, \citenamefont {Mylnikov}, \citenamefont {Bandurin},\ and\ \citenamefont {Svintsov}}]{Nikulin2021}%
  \BibitemOpen
  \bibfield  {author} {\bibinfo {author} {\bibfnamefont {Egor}\ \bibnamefont {Nikulin}}, \bibinfo {author} {\bibfnamefont {Dmitry}\ \bibnamefont {Mylnikov}}, \bibinfo {author} {\bibfnamefont {Denis}\ \bibnamefont {Bandurin}}, \ and\ \bibinfo {author} {\bibfnamefont {Dmitry}\ \bibnamefont {Svintsov}},\ }\bibfield  {title} {\enquote {\bibinfo {title} {{Edge diffraction, plasmon launching, and universal absorption enhancement in two-dimensional junctions}},}\ }\href {\doibase 10.1103/PhysRevB.103.085306} {\bibfield  {journal} {\bibinfo  {journal} {Physical Review B}\ }\textbf {\bibinfo {volume} {103}},\ \bibinfo {pages} {085306} (\bibinfo {year} {2021})}\BibitemShut {NoStop}%
\bibitem [{\citenamefont {Matov}\ \emph {et~al.}(1993)\citenamefont {Matov}, \citenamefont {Polischuk},\ and\ \citenamefont {Popov}}]{Matov1993}%
  \BibitemOpen
  \bibfield  {author} {\bibinfo {author} {\bibfnamefont {O.~R.}\ \bibnamefont {Matov}}, \bibinfo {author} {\bibfnamefont {O.~V.}\ \bibnamefont {Polischuk}}, \ and\ \bibinfo {author} {\bibfnamefont {V.~V.}\ \bibnamefont {Popov}},\ }\bibfield  {title} {\enquote {\bibinfo {title} {{Electromagnetic emission from two-dimensional plasmons in a semiconductor-dielectric structure with metal grating: Rigorous theory}},}\ }\href {\doibase 10.1007/BF02084419} {\bibfield  {journal} {\bibinfo  {journal} {International Journal of Infrared and Millimeter Waves}\ }\textbf {\bibinfo {volume} {14}},\ \bibinfo {pages} {1455--1470} (\bibinfo {year} {1993})}\BibitemShut {NoStop}%
\bibitem [{\citenamefont {Mikhailov}(1998)}]{Mikhailov1998}%
  \BibitemOpen
  \bibfield  {author} {\bibinfo {author} {\bibfnamefont {S.~A.}\ \bibnamefont {Mikhailov}},\ }\bibfield  {title} {\enquote {\bibinfo {title} {{Plasma instability and amplification of electromagnetic waves in low-dimensional electron systems}},}\ }\href {\doibase 10.1103/PhysRevB.58.1517} {\bibfield  {journal} {\bibinfo  {journal} {Physical Review B}\ }\textbf {\bibinfo {volume} {58}},\ \bibinfo {pages} {1517--1532} (\bibinfo {year} {1998})}\BibitemShut {NoStop}%
\bibitem [{\citenamefont {Zabolotnykh}\ and\ \citenamefont {Volkov}(2019)}]{Zabolotnykh_proximity}%
  \BibitemOpen
  \bibfield  {author} {\bibinfo {author} {\bibfnamefont {A.~A.}\ \bibnamefont {Zabolotnykh}}\ and\ \bibinfo {author} {\bibfnamefont {V.~A.}\ \bibnamefont {Volkov}},\ }\bibfield  {title} {\enquote {\bibinfo {title} {{Interaction of gated and ungated plasmons in two-dimensional electron systems}},}\ }\href {\doibase 10.1103/PhysRevB.99.165304} {\bibfield  {journal} {\bibinfo  {journal} {Physical Review B}\ }\textbf {\bibinfo {volume} {99}},\ \bibinfo {pages} {165304} (\bibinfo {year} {2019})}\BibitemShut {NoStop}%
\bibitem [{\citenamefont {Landau}\ \emph {et~al.}(2013)\citenamefont {Landau}, \citenamefont {Pitaevskii},\ and\ \citenamefont {Lifshitz}}]{Landau_Electrodynamics}%
  \BibitemOpen
  \bibfield  {author} {\bibinfo {author} {\bibfnamefont {Lev~Davidovich}\ \bibnamefont {Landau}}, \bibinfo {author} {\bibfnamefont {LP}~\bibnamefont {Pitaevskii}}, \ and\ \bibinfo {author} {\bibfnamefont {EM}~\bibnamefont {Lifshitz}},\ }\href@noop {} {\emph {\bibinfo {title} {Electrodynamics of continuous media}}},\ Vol.~\bibinfo {volume} {8}\ (\bibinfo  {publisher} {elsevier},\ \bibinfo {year} {2013})\ \bibinfo {note} {see ch. 3 for discussion of the electrostatic lightning rod effect and ch. 94, 95 for analytical theory of diffraction at the wedge and the metallix screen}\BibitemShut {NoStop}%
\bibitem [{\citenamefont {Noble}(1958)}]{Noble1958MethodsBO}%
  \BibitemOpen
  \bibfield  {author} {\bibinfo {author} {\bibfnamefont {Ben}\ \bibnamefont {Noble}},\ }\href@noop {} {\emph {\bibinfo {title} {Methods Based on the {W}iener-{H}opf Technique for the Solution of Partial Differential Equations}}}\ (\bibinfo  {publisher} {Pergamon Press},\ \bibinfo {year} {1958})\BibitemShut {NoStop}%
\bibitem [{\citenamefont {Cervetti}\ \emph {et~al.}(2015)\citenamefont {Cervetti}, \citenamefont {Heintze}, \citenamefont {Gorshunov}, \citenamefont {Zhukova}, \citenamefont {Lobanov}, \citenamefont {Hoyer}, \citenamefont {Burghard}, \citenamefont {Kern}, \citenamefont {Dressel},\ and\ \citenamefont {Bogani}}]{Zhukova_spectroscopy}%
  \BibitemOpen
  \bibfield  {author} {\bibinfo {author} {\bibfnamefont {Christian}\ \bibnamefont {Cervetti}}, \bibinfo {author} {\bibfnamefont {Eric}\ \bibnamefont {Heintze}}, \bibinfo {author} {\bibfnamefont {Boris}\ \bibnamefont {Gorshunov}}, \bibinfo {author} {\bibfnamefont {Elena}\ \bibnamefont {Zhukova}}, \bibinfo {author} {\bibfnamefont {Svyatoslav}\ \bibnamefont {Lobanov}}, \bibinfo {author} {\bibfnamefont {Alexander}\ \bibnamefont {Hoyer}}, \bibinfo {author} {\bibfnamefont {Marko}\ \bibnamefont {Burghard}}, \bibinfo {author} {\bibfnamefont {Klaus}\ \bibnamefont {Kern}}, \bibinfo {author} {\bibfnamefont {Martin}\ \bibnamefont {Dressel}}, \ and\ \bibinfo {author} {\bibfnamefont {Lapo}\ \bibnamefont {Bogani}},\ }\bibfield  {title} {\enquote {\bibinfo {title} {{Sub‐Terahertz Frequency‐Domain Spectroscopy Reveals Single‐Grain Mobility and Scatter Influence of Large‐Area Graphene}},}\ }\href {\doibase 10.1002/adma.201500599} {\bibfield  {journal} {\bibinfo  {journal} {Advanced Materials}\ }\textbf {\bibinfo
  {volume} {27}},\ \bibinfo {pages} {2635--2641} (\bibinfo {year} {2015})}\BibitemShut {NoStop}%
\bibitem [{\citenamefont {Toksumakov}\ \emph {et~al.}(2023)\citenamefont {Toksumakov}, \citenamefont {Ermolaev}, \citenamefont {Tatmyshevskiy}, \citenamefont {Klishin}, \citenamefont {Slavich}, \citenamefont {Begichev}, \citenamefont {Stosic}, \citenamefont {Yakubovsky}, \citenamefont {Kvashnin}, \citenamefont {Vyshnevyy} \emph {et~al.}}]{toksumakov2023anomalous}%
  \BibitemOpen
  \bibfield  {author} {\bibinfo {author} {\bibfnamefont {Adilet~N}\ \bibnamefont {Toksumakov}}, \bibinfo {author} {\bibfnamefont {Georgy~A}\ \bibnamefont {Ermolaev}}, \bibinfo {author} {\bibfnamefont {Mikhail~K}\ \bibnamefont {Tatmyshevskiy}}, \bibinfo {author} {\bibfnamefont {Yuri~A}\ \bibnamefont {Klishin}}, \bibinfo {author} {\bibfnamefont {Aleksandr~S}\ \bibnamefont {Slavich}}, \bibinfo {author} {\bibfnamefont {Ilya~V}\ \bibnamefont {Begichev}}, \bibinfo {author} {\bibfnamefont {Dusan}\ \bibnamefont {Stosic}}, \bibinfo {author} {\bibfnamefont {Dmitry~I}\ \bibnamefont {Yakubovsky}}, \bibinfo {author} {\bibfnamefont {Dmitry~G}\ \bibnamefont {Kvashnin}}, \bibinfo {author} {\bibfnamefont {Andrey~A}\ \bibnamefont {Vyshnevyy}},  \emph {et~al.},\ }\bibfield  {title} {\enquote {\bibinfo {title} {Anomalous optical response of graphene on hexagonal boron nitride substrates},}\ }\href@noop {} {\bibfield  {journal} {\bibinfo  {journal} {Communications Physics}\ }\textbf {\bibinfo {volume} {6}},\ \bibinfo {pages} {13}
  (\bibinfo {year} {2023})}\BibitemShut {NoStop}%
\bibitem [{\citenamefont {Fateev}\ \emph {et~al.}(2017)\citenamefont {Fateev}, \citenamefont {Mashinsky},\ and\ \citenamefont {Popov}}]{Fateev_rectification}%
  \BibitemOpen
  \bibfield  {author} {\bibinfo {author} {\bibfnamefont {D.~V.}\ \bibnamefont {Fateev}}, \bibinfo {author} {\bibfnamefont {K.~V.}\ \bibnamefont {Mashinsky}}, \ and\ \bibinfo {author} {\bibfnamefont {V.~V.}\ \bibnamefont {Popov}},\ }\bibfield  {title} {\enquote {\bibinfo {title} {{Terahertz plasmonic rectification in a spatially periodic graphene}},}\ }\href {\doibase 10.1063/1.4975829} {\bibfield  {journal} {\bibinfo  {journal} {Applied Physics Letters}\ }\textbf {\bibinfo {volume} {110}},\ \bibinfo {pages} {061106} (\bibinfo {year} {2017})}\BibitemShut {NoStop}%
\bibitem [{\citenamefont {Moiseenko}\ \emph {et~al.}(2024)\citenamefont {Moiseenko}, \citenamefont {Fateev},\ and\ \citenamefont {Popov}}]{moiseenko2024dissipative}%
  \BibitemOpen
  \bibfield  {author} {\bibinfo {author} {\bibfnamefont {I.M.}\ \bibnamefont {Moiseenko}}, \bibinfo {author} {\bibfnamefont {D.V.}\ \bibnamefont {Fateev}}, \ and\ \bibinfo {author} {\bibfnamefont {V.V.}\ \bibnamefont {Popov}},\ }\bibfield  {title} {\enquote {\bibinfo {title} {Dissipative drift instability of plasmons in a single-layer graphene},}\ }\href@noop {} {\bibfield  {journal} {\bibinfo  {journal} {Physical Review B}\ }\textbf {\bibinfo {volume} {109}},\ \bibinfo {pages} {L041401} (\bibinfo {year} {2024})}\BibitemShut {NoStop}%
\bibitem [{\citenamefont {Sydoruk}\ \emph {et~al.}(2012)\citenamefont {Sydoruk}, \citenamefont {Syms},\ and\ \citenamefont {Solymar}}]{Sydoruk_mirrors}%
  \BibitemOpen
  \bibfield  {author} {\bibinfo {author} {\bibfnamefont {O.}~\bibnamefont {Sydoruk}}, \bibinfo {author} {\bibfnamefont {R.~R.~A.}\ \bibnamefont {Syms}}, \ and\ \bibinfo {author} {\bibfnamefont {L.}~\bibnamefont {Solymar}},\ }\bibfield  {title} {\enquote {\bibinfo {title} {{Amplifying mirrors for terahertz plasmons}},}\ }\href {\doibase 10.1063/1.4766924} {\bibfield  {journal} {\bibinfo  {journal} {Journal of Applied Physics}\ }\textbf {\bibinfo {volume} {112}} (\bibinfo {year} {2012}),\ 10.1063/1.4766924}\BibitemShut {NoStop}%
\bibitem [{\citenamefont {Aizin}\ and\ \citenamefont {Dyer}(2012)}]{Aizin_finite}%
  \BibitemOpen
  \bibfield  {author} {\bibinfo {author} {\bibfnamefont {Gregory~R.}\ \bibnamefont {Aizin}}\ and\ \bibinfo {author} {\bibfnamefont {Gregory~C.}\ \bibnamefont {Dyer}},\ }\bibfield  {title} {\enquote {\bibinfo {title} {{Transmission line theory of collective plasma excitations in periodic two-dimensional electron systems: Finite plasmonic crystals and Tamm states}},}\ }\href {\doibase 10.1103/PhysRevB.86.235316} {\bibfield  {journal} {\bibinfo  {journal} {Physical Review B}\ }\textbf {\bibinfo {volume} {86}},\ \bibinfo {pages} {235316} (\bibinfo {year} {2012})}\BibitemShut {NoStop}%
\bibitem [{\citenamefont {Dyer}\ \emph {et~al.}(2013)\citenamefont {Dyer}, \citenamefont {Aizin}, \citenamefont {Allen}, \citenamefont {Grine}, \citenamefont {Bethke}, \citenamefont {Reno},\ and\ \citenamefont {Shaner}}]{Aizin_NPhoton}%
  \BibitemOpen
  \bibfield  {author} {\bibinfo {author} {\bibfnamefont {Gregory~C.}\ \bibnamefont {Dyer}}, \bibinfo {author} {\bibfnamefont {Gregory~R.}\ \bibnamefont {Aizin}}, \bibinfo {author} {\bibfnamefont {S.~James}\ \bibnamefont {Allen}}, \bibinfo {author} {\bibfnamefont {Albert~D.}\ \bibnamefont {Grine}}, \bibinfo {author} {\bibfnamefont {Don}\ \bibnamefont {Bethke}}, \bibinfo {author} {\bibfnamefont {John~L.}\ \bibnamefont {Reno}}, \ and\ \bibinfo {author} {\bibfnamefont {Eric~A.}\ \bibnamefont {Shaner}},\ }\bibfield  {title} {\enquote {\bibinfo {title} {{Induced transparency by coupling of Tamm and defect states in tunable terahertz plasmonic crystals}},}\ }\href {\doibase 10.1038/nphoton.2013.252} {\bibfield  {journal} {\bibinfo  {journal} {Nature Photonics}\ }\textbf {\bibinfo {volume} {7}},\ \bibinfo {pages} {925--930} (\bibinfo {year} {2013})}\BibitemShut {NoStop}%
\bibitem [{\citenamefont {Gorbenko}\ and\ \citenamefont {Kachorovskii}(2024)}]{Gorbenko_LateralPC}%
  \BibitemOpen
  \bibfield  {author} {\bibinfo {author} {\bibfnamefont {Ilya}\ \bibnamefont {Gorbenko}}\ and\ \bibinfo {author} {\bibfnamefont {Valentin}\ \bibnamefont {Kachorovskii}},\ }\bibfield  {title} {\enquote {\bibinfo {title} {Lateral plasmonic crystals: Tunability, dark modes, and weak-to-strong coupling transition},}\ }\href {\doibase 10.1103/PhysRevB.110.155406} {\bibfield  {journal} {\bibinfo  {journal} {Phys. Rev. B}\ }\textbf {\bibinfo {volume} {110}},\ \bibinfo {pages} {155406} (\bibinfo {year} {2024})}\BibitemShut {NoStop}%
\end{thebibliography}%

\maketitle

\end{document}